\documentclass{emulateapj}
\usepackage{color,amsmath,graphics,graphicx,natbib}
\begin{document}

\title{The Accretion History of AGN: A Newly Defined Population of Cold Quasars}
\author{Allison Kirkpatrick\altaffilmark{1}, C. Megan Urry\altaffilmark{2,3}, Jason Brewster\altaffilmark{4}, Kevin C. Cooke\altaffilmark{5}, Michael Estrada\altaffilmark{1}, Eilat Glikman\altaffilmark{6}, Kurt Hamblin\altaffilmark{4}, Tonima Tasnim Ananna\altaffilmark{2,3}, Casey Carlile\altaffilmark{1}, Brandon Coleman\altaffilmark{1}, Jordan Johnson\altaffilmark{1}, Jeyhan S. Kartaltepe\altaffilmark{5}, Stephanie M. LaMassa\altaffilmark{7}, Stefano Marchesi\altaffilmark{8}, Meredith Powell\altaffilmark{2,3,9}, Dave Sanders\altaffilmark{10}, Ezequiel Treister\altaffilmark{11}, Tracey Jane Turner\altaffilmark{4,12}}

\altaffiltext{1}{Department of Physics \& Astronomy, University of Kansas, Lawrence, KS 66045, USA, akirkpatrick@ku.edu}
\altaffiltext{2}{Yale Center for Astronomy \& Astrophysics, New Haven, CT 06520, USA}
\altaffiltext{3}{Department of Physics, Yale University, PO BOX 201820, New Haven, CT 06520, USA}
\altaffiltext{4}{Department of Physics, University of Maryland Baltimore County, Baltimore, MD 21250, USA}
\altaffiltext{5}{School of Physics and Astronomy, Rochester Institute of Technology, Rochester, NY 14623, USA}
\altaffiltext{6}{Department of Physics, Middlebury College, Middlebury, VT 05753, USA}
\altaffiltext{7}{Space Telescope Science Institute, 3700 San Martin Dr, Baltimore, MD 21218, USA}
\altaffiltext{8}{INAF - Osservatorio di Astrofisica e Scienza dello Spazio di Bologna, Via Piero Gobetti, 93/3, 40129, Bologna, Italy}
\altaffiltext{9}{Kavli Institute for Particle Astrophysics and Cosmology, Stanford University, 452 Lomita Mall, Stanford, CA 94305, USA}
\altaffiltext{10}{Institute for Astronomy, University of Hawaii, 2680 Woodlawn Drive, Honolulu, HI 96822, USA}
\altaffiltext{11}{Instituto de Astrof{\'i}sica, Facultad de F{\'i}sica, Pontificia Universidad Cat{\'o}lica de Chile, Casilla 306, Santiago 22, Chile}
\altaffiltext{12}{Center for Space Science and Technology, University of Maryland Baltimore County, 1000 Hilltop Circle, Baltimore, MD 21250, USA}

\begin{abstract}
Quasars are the most luminous of active galactic nuclei (AGN), and are perhaps responsible for quenching star formation in their hosts. The Stripe 82X catalog covers 31.3 deg$^2$ of the Stripe 82 field, of which the 15.6 deg$^2$ covered with {\it XMM-Newton} is also covered by {\it Herschel}/SPIRE. We have 2500 X-ray detected sources with multi-wavelength counterparts, and 30\% of these are unobscured quasars, with $L_X > 10^{44}\,$erg/s and $M_B < -23$. We define a new population of quasars which are unobscured, have X-ray luminosities in excess of $10^{44}\,$erg/s, have broad emission lines, and yet are also bright in the far-infrared, with a 250\,$\mu$m flux density of $S_{\rm 250}>30\,$mJy. We refer to these {\it Herschel}-detected, unobscured quasars as  ``Cold Quasars''. A mere 4\% (21) of the X-ray- and optically-selected unobscured quasars in Stripe 82X are detected at 250\,$\mu$m.
These Cold Quasars lie at $z\sim1-3$, have $L_{\rm IR}>10^{12}\,L_\odot$, and have star formation rates of $\sim200-1400\,M_\odot$/yr. Cold Quasars are bluer in the mid-IR than the full quasar population, and 72\% of our Cold Quasars have {\it WISE} W3 $<$ 11.5 [Vega], while only 19\% of the full quasar sample meets this criteria. Crucially, Cold Quasars have on average
$\sim9\times$ as much star formation as the main sequence of star forming galaxies at similar redshifts. Although dust-rich, unobscured quasars have occasionally been noted in the literature before, we argue that they should be considered as a separate class of quasars due to their high star formation rates. This phase is likely short-lived, as the central engine and immense star formation consume the gas reservoir. Cold Quasars are type-1 blue quasars that reside in starburst galaxies.
\end{abstract}

\section{Introduction}
The origin of luminous quasars is perhaps a dramatic story, wherein two immense galaxies collide, fueling a burst of star formation and triggering rapid growth of a supermassive black hole \citep[e.g.,][]{hopkins2006}. In the major merger paradigm,
the active galactic nucleus (AGN) goes through an obscured growth phase, where the accretion disk is hidden by dust, while the host galaxy experiences a period of enhanced star formation. This phase ends when the AGN launches winds powerful enough to blow away some of the 
circumnuclear obscuring dust so that the accretion disk becomes visible in the optical as star formation quenches in the host galaxy \citep{glikman2004,glikman2012,pontzen2012}. Quasars are the most luminous of AGN, and it has been suggested that they are almost universally produced by major mergers \citep{treister2012}.

AGN can heat torus and circumnuclear dust to $>$1000\,K, producing IR emission from $\lambda = 1-40\mu$m \citep[e.g.,][]{elvis1994,netzer2007,mullaney2011}.
Additionally, AGN heat and excite physically extended gas clouds to form the narrow line region, which in extreme cases of merging galaxies can be larger than the galaxy itself \citep[e.g.,][]{hainline2016}.
The narrow line region contains gas ionized by the radiation field of the AGN, and it also contains dust \citep{groves2006}. The emission from the AGN torus or circumnuclear environment 
typically peaks around 25\,$\mu$m, although the exact temperature depends on the incident radiation field and 
optical depth. In AGN where the central nucleus has a lot of obscuring material, galactic dust can be 
even colder, peaking at wavelengths longer than 25\,$\mu$m, as the incident radiation field is absorbed and re-radiated through large column densities. It is 
possible, in heavily obscured cases, that the AGN can account for all of the far-IR emission \citep{sajina2012,ricci2017}. Such heavily obscured objects are thought to be rare, as the 
amount of obscuration increases with stellar mass \citep{buchner2017a,buchner2017b}. In the major merger paradigm, heavily obscured phases are likely short lived (20\% of the blue quasar phase), occurring briefly as the quasar is in the transitional blowout phase, after which it becomes optically luminous \citep{glikman2012}. 
The quasar lifetime is only weakly constrained through clustering measurements to be $\sim10^6-10^9$ yr \citep{conroy2013,laplante2016}. Further complicating matters, the IR emission in unobscured quasars can vary dramatically from source to source \citep{lyu2018}. Dust can arise from the torus, narrow-line region \citep{groves2006,mor2009}, or AGN outflows \citep{murray2005}, and the shape of the IR emission depends sensitively on viewing angle and dust covering fraction. Different models also use smaller \citep[$a=0.005-0.25\mu$m][]{fritz2006} or larger \citep[$a=0.1-1\,\mu$m][]{honig2010} grain sizes and compositions. These assumptions can change the shape of the IR emission, particularly around the silicate absorption lines. However, \citet{gonzalez2019} demonstrate that IR photometry can only distinguish between any models when there is less than 50\% contamination from the host galaxy. This is only the case for the nearest sources.

The location of the obscuration in quasars produced by mergers is difficult to tease out, as it can in principal arise from any dust or gas along the line of sight. The traditional torus model places the obscurer in a compact, donut-like structure within a few parsecs of the accretion disk. In the most successful torus models, the torus does not have sharp edges, is physically related to the accretion disk, and is made of clumpy clouds of dust and gas \citep[e.g.,][]{elitzur2006}. Obscured AGN can be the product of a viewing angle through the thicker parts of the torus \citep{urry1995}. The torus is structurally related to the supermassive black hole by being under its gravitational influence, absorbing radiation from the accretion disk, and possibly being formed from winds off of the accretion disk. The host galaxy itself is another source of obscuration. Gas and dust in the host galaxy are not associated with the AGN structure, even if they are circumnuclear. \citet{buchner2017a} and \citet{buchner2017b} measured hydrogen column densities from the host galaxy and the AGN and found that the galaxy can account for
Compton thin obscuration ($N_H = 10^{22}$\,cm$^{-2}$) but not Compton thick ($N_H> 10^{24}$\,cm$^{-2}$). Dust absorption is commonly parameterized by the optical depth of the 9.7\,$\mu$m silicate absorption feature \citep{spoon2007}. High silicate optical depth is observed in obscured AGN, and this dust may arise from the torus \citep{hatz2015}. In local mergers, \citet{goulding2012} showed that silicate dust absorption correlates with the viewing angle of the host galaxy and not with gas column densities of the AGN measured from the X-ray. Dust and gas obscuration can arise from a foreground screen rather than the AGN structure itself \citep{hatz2015,buchner2017a}. In fact, mergers in particular exhibit a clear correlation between gas and dust obscuration and merger stage, with late stage mergers being more obscured \citep{sanders1996,hopkins2008,petric2011,stierwalt2013,ricci2017}.

Quasars may substantially affect the star formation of their host galaxies through feedback processes, though the exact nature of that role remains unclear. In AGN, winds can either blow out the interstellar medium and quench star formation \citep[negative feedback;][]{hopkins2006}, or they can compress the interstellar medium and trigger star formation \citep[positive feedback;][]{ciotti2007}. Accretion onto the most massive black holes is observed to produce collimated jets that sweep through the galaxy producing kinetic feedback \citep{heckman2014}. However, in a limited number of sources, star formation is observed to be enhanced in X-ray-selected and optically-selected AGN \citep{salome2015,mahoro2017,perna2018}. Further complicating matters, other studies indicate no link between AGN activity and either enhanced or decreased star formation. For example, \citet{stanley2017} find no decrease in star formation rate (SFR) with AGN bolometric luminosity for the optically-selected AGN from the Sloan Digital Sky Survey (SDSS). \citet{omont2003} observes $L_{\rm IR}\sim10^{13}\,L_\odot$, indicating star formation rates $>1000\,M_\odot$/yr in quasars at $z\sim2$. Recently, \citet{schulze2019} found that 20 unobscured quasars at $z\sim2$ have no statistically significant difference in SFRs than star-forming galaxies, based on ALMA continuum measurements. In a small local sample, unobscured quasars were shown to have similar dust masses as obscured quasars, indicating that unobscured quasars might not be in a later evolutionary stage \citep{shangguan2019}.

One problem in synthesizing literature results into a coherent picture is that selection in different wavelength regimes can produce biased samples \citep{hickox2009,azadi2017}, particularly when studying obscured AGN. 
As 75\% of AGN growth is obscured \citep{treister2004,ananna2019}, these sources provide the most insight into black hole---galaxy coevolution. Unobscured AGN, and in particular quasars, should provide the clearest insight into how massive galaxies die, as unobscured quasars have presumably blown out their own obscuring material. Their immense energies should start to affect their hosts and usher in the passive galaxy phase. But making this connection directly is hard. Precisely because unobscured quasars are so luminous, it becomes very difficult to study their host galaxy properties. Quasars outshine their hosts in the optical and mid-IR photometry \citep{hainline2011,stern2012}. They contaminate the optical emission lines used to study star formation, particularly $H_\alpha$ \citep{baldwin1981}. Submillimeter emission lines, which trace the dense interstellar medium (ISM) in star-forming regions, are also enhanced in AGN \citep{imanishi2016,kamenetzky2016,kirkpatrick2019}. The most robust tracer of the host galaxy is therefore the far-IR, as AGN are expected not to be able to heat dust beyond $\lambda>100\,\mu$m, as is evidenced by the sharp drop in far-IR emission seen in X-ray luminous AGN \citep{mullaney2011,rosario2012}. \citet{lyu2017} argue that AGN in fact cannot account for much emission at wavelengths longer than 20\,$\mu$m, simply due to energy balance arguments. 

In this paper, we present an unbiased look at the interstellar medium of unobscured quasars.
This is the first in a series of papers examining in detail the multi-wavelength emission of X-ray selected AGN with far-IR emission in the Accretion History of AGN survey (AHA\footnote{\url{http://project.ifa.hawaii.edu/aha/}}; PI, M. Urry). We discuss the survey and infrared sample selection in Section \ref{sec:survey}, including our definition of Cold Quasars. We discuss in Section \ref{sec:IR} how the IR emission of Cold Quasars compares with unobscured quasars. Our conclusions are listed in Section \ref{sec:conc}.
Throughout this paper, we assume a standard cosmology with $H_{0}=70\,\rm{km}\,\rm{s}^{-1}\,\rm{Mpc}^{-1}$, $\Omega_{\rm{M}}=0.3$ and $\Omega_{\Lambda}=0.7$. 

\section{A Multi-wavelength AGN Survey}
\label{sec:survey}
Large volumes of the universe must be surveyed in order to discover a representative sample of rare objects like high-luminosity AGN.
This paper is part of AHA, which is a multi-wavelength survey that assembles data from the Stripe 82X \citep[31\,deg$^2$;][]{lamassa2013a,lamassa2013b,lamassa2015}, COSMOS \citep[2\,deg$^2$;][]{cappelluti2007,scoville2007,elvis2009,civano2016,marchesi2016}, and GOODS-South fields \citep[0.1\,deg$^2$;][]{brandt2001,giacconi2002,giavalisco2004}.
These fields comprise a
``wedding cake'' tiered X-ray survey, where each cake layer probes a different area and flux limit, and thus a
different luminosity-redshift range. Stripe 82X covers the most luminous sources, while COSMOS and GOODS-South contain less luminous and/or more obscured AGN. We focus here on the Stripe 82X component, which has the most luminous X-ray sources in our survey and will discuss the IR AGN in the full AHA survey in a subsequent paper.

Our team surveyed 4.6 deg$^2$ and 15.6 deg$^2$ of the Sloan Digital Sky Survey (SDSS) field Stripe 82 with {\it XMM-Newton} in AO10 and AO13 \citep{lamassa2013,lamassa2016}. We combined these observations with a further 5.6 deg$^2$ of {\it XMM-Newton} archival pointings and 6.0 deg$^2$ of {\it Chandra} archival pointings in Stripe 82 for a total Stripe 82X survey area of 31.3 deg$^2$.
From this catalog, 15.6 deg$^2$, covered with {\it XMM-Newton} in AO13, is also fully covered by {\it Herschel Space Observatory} observations with the SPIRE instrument as part of the {\it Herschel} Stripe82 Survey \citep[HerS;][]{viero2014}. AO13 has coadded exposure times of 6--8 ks \citep{lamassa2016}. Additionally, there are 125 sources from AO10 that fall in the HerS survey region \citep{lamassa2013}. We refer the reader to Figure 1 in \citet{ananna2017} for the layout of Stripe 82 covered by {\it XMM} and {\it Herschel}. 
Multi-wavelength counterpart matching was done in \citet{lamassa2016} and \citet{ananna2017}. The authors use the co-added SDSS catalogs \citep{jiang2014,fliri2016}, which reach 2.5 mag deeper than single epoch data, to increase the likelihood of each X-ray source having an optical counterpart. The authors match the X-ray positions to multi-wavelength counterparts spanning the UV to the mid-IR. This portion of Stripe 82 is covered by warm {\it Spitzer} (3.6 \& 4.5\,$\mu$m only) through the {\it Spitzer}/HETDEX Exploratory Large Area (SHELA) survey \citep{papovich2016}, the {\it Spitzer} IRAC Equatorial Survey \citep[SpIES;][]{timlin2016}, and by {\it WISE} in the mid-IR. It is also covered in the near-IR by the Vista Hemisphere Survey \citep{mcmahon2013} and the UKIRT Infrared Deep Sky Survey \citep{lawrence2007}, and in the UV by {\it GALEX}. Full details of the observations and counterpart identification can be found in \citet{lamassa2016} and \citet{ananna2017}. The 16\,deg$^2$ portion of Stripe 82X that overlaps with the HerS survey area contains 3200 X-ray sources \citep{lamassa2016}. We remove 295 sources that do not have robust multi-wavelength counterparts (and hence photo-$z$s), as indicated by a counterpart quality flag of 3 or 4 in the \citet{ananna2017} photometric catalog. This leaves 2905 X-ray sources.

\subsection{The Herschel/X-ray Subsample}
\begin{figure}
\centering
\includegraphics[width=3in]{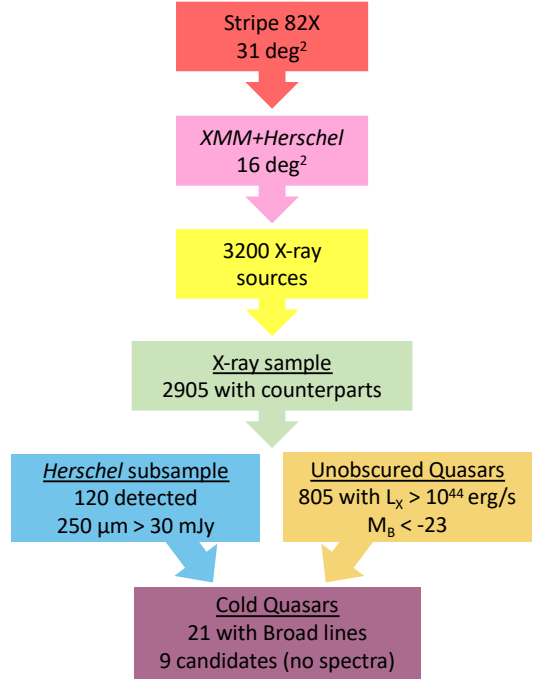}
\caption{Visual summary of our main sample selection criteria. \label{flow}.}
\end{figure}

One of the goals of AHA is to determine the host galaxy properties of the most luminous X-ray sources. This requires unambiguous observations of the host galaxy, which, for unobscured AGN, can best be done at longer wavelengths, such as the far-IR and sub-mm. A luminous AGN can contaminate all other wavelengths. 
Of the 2905 X-ray sources, 120 galaxies are also detected at 250\,$\mu$m, which we refer to as the {\it Herschel} subsample. The HerS survey has a $3\sigma$ detection limit of $S_{250\,\mu{\rm m}}=30\,$mJy \citep{viero2014}. In a purely star-forming galaxy, this flux detection limit corresponds to an SFR of 174\,$M_\odot$/yr at $z=1$ and 758\,$M_\odot$/yr at $z=2$, estimated using the \citet{kirkpatrick2012} templates. We summarize the sample selection Figure \ref{flow}.

\begin{figure}
\includegraphics[width=3.3in]{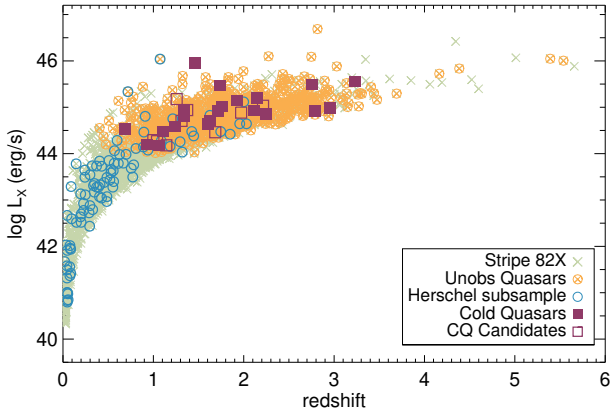}
\caption{The X-ray ($0.5-10$\,kev) luminosity and redshift distribution of the full Stripe 82X sample in the 16 deg$^2$ region of the sky covered by both {\it XMM-Newton} and {\it Herschel} (light green crosses; 2905 galaxies). The 120 galaxies detected in both the X-ray and at 250\,$\mu$m form our {\it Herschel} subsample (blue circles). We refer to the X-ray sources with $L_X>1\times10^{44}$\,erg/s and $M_B<-23$ as unobscured quasars (orange circles with crosses). The unobscured quasars with a 250\,$\mu$m detection are our Cold Quasar subsample (purple filled squares). We identify Cold Quasar candidates as sources that meet the stated definitions but lack optical broad-line emission, either due to not having a spectrum or having a spectrum with a low signal-to-noise ratio.\label{x_z}}
\end{figure}

Ninety-nine galaxies from the {\it Herschel} subsample have spectroscopic redshifts measured from SDSS spectra (from DR12, DR13, or DR14). The remaining 21 have photometric redshifts determined by fitting the UV/optical/NIR data with the LePhare code \citep{arnouts1999,ilbert2006}. 
We show the full band X-ray luminosity (throughout the text, $L_X = L_{0.5-10\,{\rm kev}}$) and redshift distribution of the Stripe 82X sources and {\it Herschel} subsample in Figure \ref{x_z}. The X-ray luminosities have been $k$-corrected using $\Gamma = 2.0$ for the hard band and $\Gamma = 1.8$ for the soft band \citep{lamassa2016}, but they are not corrected for absorption.
 
\subsection{Cold Quasar Definition}
\label{sec:cq}
We find that many of our {\it Herschel} subsample are unobscured, luminous AGN. We call unobscured quasars any source with $L_X>1\times10^{44}$\,erg/s and $M_B <-23$ \citep{schmidt1983,mcdowell1989}. To transform between the SDSS filters and the B band, we used the equation 
\begin{equation}
 m_B=m_g + 0.17(m_u-m_g)+0.11
\end{equation}
\citep{jester2005}. From the full X-ray sample, 805 sources meet this definition. We refer to these as the unobscured quasars throughout the text and figures.
Of the {\it Herschel} subsample, 32 galaxies satisfy these criteria, 23 of which have optical broad lines, confirming that they are type-1 quasars. Two of these sources are blazars, based on their reported radio flux densities in the Very Large Array's FIRST survey \citep{white1997}. These two sources also have 500\,$\mu$m emission that is brighter than the 250\,$\mu$m emission. This ``rising'' emission indicates likely contamination of the far-IR emission from non-thermal sources. We remove these two quasars from the sample. Herein, we create a new definition to categorize quasars. We refer to the remaining 21 galaxies as the ``Cold Quasars''.
The remaining nine {\it Herschel} sources either do not have an SDSS spectrum or have a spectrum with low signal to noise ratio, making it difficult to distinguish broad lines. We refer to these nine galaxies as Cold Quasar Candidates throughout the text. We will analyze all optical spectra from the full {\it Herschel} subsample in a future paper (Estrada et al., in prep). 

\begin{figure*}
\centering
\includegraphics[width=2in]{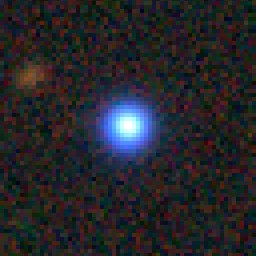}
\includegraphics[width=2.5in]{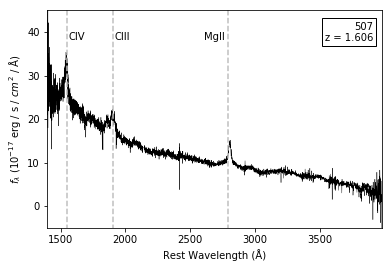}
\caption{An example of a Cold Quasar (ID 507, $z=1.606$). {\it Left--}An RGB image, where blue is the $g$-band filter from the DECam Legacy Survey, green is the $r$-band, and red is the $z$-band. Each side of the image is 20''. The blue color indicates that this quasar is not dust reddened. All of our Cold Quasars are blue point sources. {\it Right--}Rest frame optical SDSS spectrum with the broad-lines labeled. The continuum is blue two broad-lines are evident. Spectra and images for all {\it Herschel} sources are available at https://kirkpatrick.ku.edu/CQ/. \label{visual}}
\end{figure*}

\begin{figure*}
\centering
\includegraphics[width=2.3in]{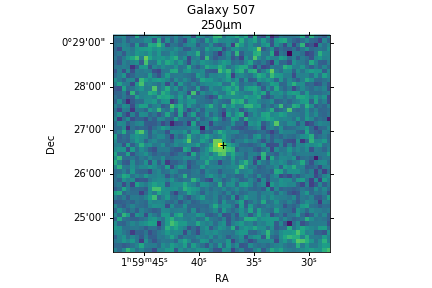}
\includegraphics[width=2.3in]{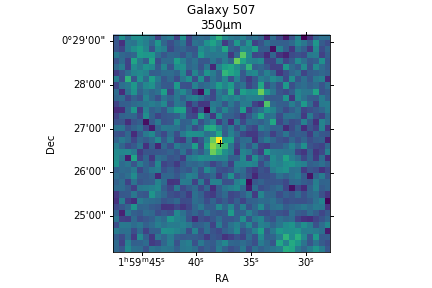}
\includegraphics[width=2.3in]{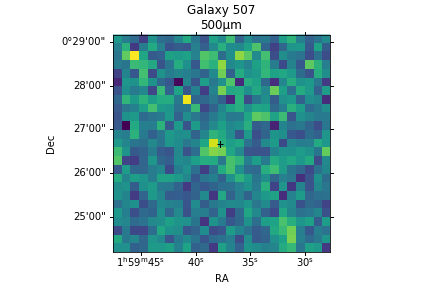}
\caption{The far-IR emission of Cold Quasar 507 ($z=1.606$). From top to bottom, we show the 250\,$\mu$m, 350\,$\mu$m, and 500\,$\mu$m {\it Herschel}/SPIRE images, with the X-ray position marked with the cross. \label{IR}}
\end{figure*}

All Cold Quasars resemble point sources in the optical and are unresolved in the {\it Herschel} bands (Figures \ref{visual} and \ref{IR}). We show all optical and {\it Herschel} images on our website (https://kirkpatrick.ku.edu/CQ/). \footnote{We use the DESI Legacy Imaging Surveys \citep{dey2019} to create optical images from the $grz$ bands of our Cold Quasars. Our sources lie in the footprint covered by the Dark Energy Camera Legacy Survey \citep{flaugher2015}, observed by the Dark Energy Camera on the 4-m Blanco telescope at the Cerro Tololo Inter-American Observatory. The Legacy Surveys reach a $z$-band limiting magnitude of 22.5 mag, making them deeper than the images in the SDSS Data Releases. However, we do not use the photometry from the Legacy Surveys for any spectral fitting or analysis.} They lie between $0.5<z<3.5$, due to the extreme X-ray luminosities. We list the properties of the Cold Quasars in Table \ref{properties}. All Cold Quasars are detected in all three {\it Herschel}/SPIRE bands.

Red Quasars are type-1 quasars that have a red continuum in the optical, indicating a moderate amount of dust ($E(B-V)>0.25$) along the line of sight that does not fully obscure the broad-line region \citep{urrutia2008,glikman2013,glikman2018}. These quasars are among the most luminous objects in the universe and are generally found in late-stage mergers \citep{glikman2015}. Red Quasar candidates can be selected using $r_{\rm AB}-K_{\rm Vega} > 5\,\wedge\,J-K>1.5$ \citep{glikman2013}. According to this color selection, none of our Cold Quasars or Cold Quasar candidates are Red Quasars. We find the Cold Quasar population is distinct from the Red Quasar population in that they do not have reddened continua. 

\begin{deluxetable*}{l cc l l  l l l  l c}
\tablecolumns{11}
\tablecaption{Cold Quasar Properties \label{properties}}

\tablehead{ \colhead{ID\tablenotemark{a}} & \colhead{RA (J2000)} & \colhead{Dec (J2000)} &  \colhead{$z$} & \colhead{$\log L_{0.5-10 \rm{keV}}$} & 
\colhead{$\log L_{\rm bol}$} &
\colhead{$\log L_{\rm IR}$} & \colhead{$f_{\rm AGN}$} & \colhead{$\log M_\ast$} & \colhead{SFR} \\
\colhead{} & \colhead{} & \colhead{} &  \colhead{} & \colhead{(erg/s)} & \colhead{(erg/s)} & \colhead{$(L_\odot)$} & \colhead{$(8-1000\mu$m)} & \colhead{$(M_\odot)$} & \colhead{($M_\odot$/yr)}}

\startdata
 475  	&  01:19:48.4		& +00:43:54.28  	& 1.754  & 45.01	& 46.85$\pm$0.01 &  13.41$\pm$0.05 & 0.86$\pm$0.13 & 12.00$\pm$0.12	& 588$\pm$112  \\
 507		& 01:59:37.8		&  +00:26:39.91	& 1.606	& 44.64	& 46.60$\pm$0.01 & 13.07$\pm$0.03 & 0.56$\pm$0.04 & 10.80$\pm$0.13	& 835$\pm$102 \\
 2435 	& 00:57:43.7		& $-$00:11:57.92	& 1.067	& 44.19	& 45.78$\pm$0.01 & 12.69$\pm$0.03 & 0.21$\pm$0.02 & 10.21$\pm$0.78	& 616$\pm$56 \\
 2480 	& 00:58:28.9		& +00:13:45.25		& 1.239	& 44.59	& 45.87$\pm$0.01 & 12.53$\pm$0.07 & 0.55$\pm$0.09 & 10.69$\pm$0.11	& 240$\pm$79 \\
2551 	& 00:59:46.6		& +00:02:54.61 	& 0.682	& 44.54	& 45.63$\pm$0.03 & 12.41$\pm$0.04 &  0.45$\pm$0.04 & 10.89$\pm$0.10	& 226$\pm$36\\
2651  	&  01:01:13.3  		& $-$00:29:45.20 	& 1.337  & 44.80	& 46.11$\pm$0.01 & 12.60$\pm$0.06  & 0.52$\pm$0.08 & 10.98$\pm$0.16	& 306$\pm$82 \\
3122  	&  01:09:01.0  		& +00:01:37.44  	& 2.955  &  44.98	& 46.78$\pm$0.20\tablenotemark{b} & 13.51$\pm$0.04  & 0.73$\pm$0.07 &  \nodata	& 1393$\pm$299 \\
3716  	&  01:19:00.4  		& $-$00:01:57.68 	& 2.750  &  45.50	& 47.85$\pm$0.20\tablenotemark{b} & 13.48$\pm$0.03  & 0.79$\pm$0.05& \nodata 	& 976$\pm$243 \\
3819  	&  01:21:10.0  		& $-$00:29:10.36 	& 1.719  &  44.92	& 46.25$\pm$0.01 &12.67$\pm$0.07 & 0.43$\pm$0.07 & 10.90$\pm$0.15	& 428$\pm$117 \\
3871  	&  01:22:49.7  		& $-$00:07:07.13	& 1.631  &  44.71	& 46.14$\pm$0.01 & 12.64$\pm$0.16  & 0.48$\pm$0.18 & 11.27$\pm$0.16 & 358$\pm$251  \\
4077  	&  01:30:34.0  		& $-$00:21:06.61 	& 3.234  &  45.55	& 47.95$\pm$0.20\tablenotemark{b} & 13.15$\pm$0.10  & 0.36$\pm$0.09 & \nodata	& 1435$\pm$511 \\
4252     	&  01:35:54.4  		& $-$00:22:31.91 	& 2.249  &  44.85 	& 46.46$\pm$0.03 & 13.01$\pm$0.06  & 0.34$\pm$0.05 & 10.71$\pm$0.62 & 1062$\pm$197 \\
4285		 & 01:37:26.4	  	& +00:11:52.45	& 1.103	& 44.47	& 46.20$\pm$0.00 & 12.64$\pm$0.05 &  0.60$\pm$0.07 & 10.84$\pm$0.14	& 276$\pm$73 \\
4324  	&  01:38:14.7  		&  +00:00:05.78   &  2.144  & 45.20 & 46.92$\pm$0.01 & 13.35$\pm$0.07 &  0.65$\pm$0.18 & 11.43$\pm$0.13	& 1255$\pm$153  \\
4336  	&  01:38:25.3  	& $-$00:05:34.39 &  1.341  & 44.94	& 46.73$\pm$0.00 & 12.95$\pm$0.03 & 0.78$\pm$0.05 & 11.17$\pm$0.10	& 315$\pm$79  \\
4472  	&  01:40:33.8  	&  +00:02:30.05   &  1.921  & 45.15 	& 46.19$\pm$0.04 & 13.06$\pm$0.04  & 0.46$\pm$0.04	& 10.45$\pm$0.19 & 995$\pm$150 \\
4668  	&  01:43:01.9  	& $-$00:26:56.54 &  2.786  & 44.93 & 47.01$\pm$0.01 & 13.32$\pm$0.06 & 0.68$\pm$0.09 & 11.20$\pm$0.12	& 1070$\pm$416 \\
4979	 	& 01:08:21.1	 	& $-$00:02:28.51 	& 0.930	& 44.19	& 46.20$\pm$0.00 &  12.44$\pm$0.06  & 0.63$\pm$0.09  & 10.52$\pm$0.14  & 162$\pm$59 \\ 
5074  	&  01:49:40.2  	&  +00:17:17.76  	& 1.464  & 45.95 	& 46.73$\pm$0.00 & 12.88$\pm$0.03  & 0.74$\pm$0.06 & 11.05$\pm$0.12	& 309$\pm$87 \\
5097  	&  01:49:58.3  	& $-$00:30:25.00 &  2.111  & 44.94	& 46.37$\pm$0.03 & 13.22$\pm$0.08 & 0.64$\pm$0.21 & 10.98$\pm$0.16	& 949$\pm$145  \\
5122  	&  01:50:34.5  	& $-$00:02:00.46 &  1.740  & 45.47	& 46.54$\pm$0.01 & 13.29$\pm$0.05  & 0.76$\pm$0.13 & 10.31$\pm$0.86	& 725$\pm$124
\enddata
\tablenotetext{a}{The ID is the Rec\_No from \citet{lamassa2016}.}
\tablenotetext{b}{The bolometric luminosity is determined by Equation \ref{eqbol}.}
\end{deluxetable*}

\subsection{Host Galaxy Properties}
\label{fitmethod}
{We determined the IR luminosities and IR AGN contribution of all {\it Herschel} X-ray sources by decomposing the WISE and {\it Herschel} photometry into an AGN and host galaxy component. We use the Featureless AGN template from \citet{kirkpatrick2012} and three possible host templates: the $z\sim1$ Star Forming template from \citet{kirkpatrick2012}, the submillimeter galaxy template from \citet{pope2008}, and the SB5 template from \citet{mullaney2011}. These three templates were chosen to cover a range in dust temperatures and far-IR to mid-IR emission. The Featureless AGN template is derived from the emission of X-ray luminous AGN at $z\sim1-3$. Due to the scarcity of data points in the IR, we fit the simplest model possible: $model = N_1 \times AGN + N_2 \times Host$, where we allow the normalizations, $N_1$ and $N_2$, to vary simultaneously. We use a $\chi^2$ minimization technique to determine the best fit. Since the photometric uncertainty is large in the mid- and far-IR, we use a Monte Carlo simulation to determine the uncertainties of the model. We randomly
sample each data point within its error and refit the model 100 times. The final normalizations are the median of these 100 trials, and the associated uncertainties are the standard deviations. All of the Cold Quasars are best fit using the submillimeter galaxy template, which has the largest far-IR to mid-IR ratio and is empirically derived from starbursts at $z=1-2$. For four Cold Quasars, we also have to vary the slope of the Featureless AGN template to be steeper, indicating more obscuration in the torus. We show all the fits in the Appendix. 

The Featureless AGN template allows for a large contribution of the far-IR to be from the AGN itself (this template peaks around $\lambda\sim30\mu$m). This amount of far-IR heating was determined empirically \citep{kirkpatrick2012}. Without more data covering the crucial transition range between AGN- and host-dominated emission ($\lambda\sim10-70\mu$m), it is impossible to further constrain how much of the far-IR originates with the AGN heating. The Featureless AGN template accounts for more of the far-IR heating than standard QSO templates \citep{elvis1994,lyu2017}. We test how much the $L_{\rm IR}$ and SFR changes when we use the \citet{elvis1994} Type-1 QSO template, and we find the change to be negligible.

We experimented with more complex fitting methods such as {\tt SED3FIT} \citep{berta2013} and {\tt CIGALE} \citep{burgarella2005,boquien2019}. The SFRs were generally a factor of $2-3\times$ higher than with our fitting method, and the AGN contribution to the IR was found to be much lower. This is due to the fact that the AGN templates used in both codes fall off around 10\,$\mu$m, so they account for a negligible amount of the far-IR heating. For $>60\%$ of the sample, we found the UV-optical fitting to be unreliable. Both {\tt CIGALE} and {\tt SED3FIT} determined the UV-Optical was dominated by galaxy emission, despite spectral evidence to the contrary. We verified that the optical emission of our Cold Quasars can be completely accounted for by an unobscured quasar, by scaling the optical luminous QSO template from \citet{richards2006}. All of the Cold Quasars can be fit with this template, and only 4 require any extinction, which we model with an SMC-like dust curve \citep{draine2003}. For these four quasars, E(B$-$V)$ <0.2$. When using {\tt SED3FIT} or {\tt CIGALE}, the resulting SFRs are high due to the assumed contribution from unobscured stars in the UV and the slope of the extrapolated longer wavelength emission ($\lambda>300\,\mu$m). Due to these issues, we opted not to use the {\tt SED3FIT} or {\tt CIGALE} results. Exploring these discrepancies in more detail will be the subject of a future paper.

We calculate $f_{\rm AGN}$ by integrating under the Featureless AGN template to obtain the $L_{\rm IR}$ of the AGN component, and we express this as a fraction of the total $L_{\rm IR} (8-1000\,\mu$m). To determine the SFR, we integrate the host galaxy template from $8-1000\,\mu$m and then use the
relationship from \citet{murphy2011}:
\begin{equation}
{\rm SFR}\,[M_\odot/{\rm yr}] = (1.59\times10^{-10}) \times L_{\rm IR}^{\rm host}\, [L_\odot]
\end{equation}

In unobscured quasars, the AGN dominates the emission at nearly all wavelengths, making stellar mass notoriously difficult to determine. We estimate $M_\ast$ from the black hole mass. We obtain $M_\bullet$ for 18 of our Cold Quasars from the SDSS DR7 spectral catalogue published in \citet{shen2011}, which includes $L_{\rm bol}$ estimated from the 5100\,\AA, 3000\,\AA, 1350\,\AA\ monochromatic luminosity and $M_\bullet$ estimated from H$\beta$, C {\sc iv}, or Mg {\sc ii}. We use the relation from \citet{bennert2011} to calculate $M_\ast$:
\begin{align}
\log \frac{M_\bullet}{10^8M_\odot}&=1.12 \log \frac{M_\ast}{10^{10}M_\odot}
+(1.15\pm0.15)\log(1+z) \nonumber\\
&-0.68+(0.16\pm0.06)
\end{align}
We adopt the $L_{\rm bol}$ listed in \citet{shen2011}. For the three galaxies not included in this catalog, we calculated $L_{\rm bol}$ from $L_X$ using the following relation from \citet{lusso2012}:
\begin{align}
\label{eqbol}
\log \frac{L_{\rm Bol}}{L_X} = -0.020x^3+0.114x^2+0.310x+1.296
\end{align}
where $x=\log L_{\rm Bol} -12$ in solar luminosity units.
We list the host galaxy properties IR luminosity ($L_{\rm IR} [8-1000\,\mu{\rm m}]$), stellar mass ($M_\ast$), $f_{\rm AGN}$, and SFR in Table \ref{properties}.}

\section{Results \& Discussion}
\label{sec:IR}
\subsection{Infrared Colors}
\begin{figure}
\includegraphics[width=3.3in]{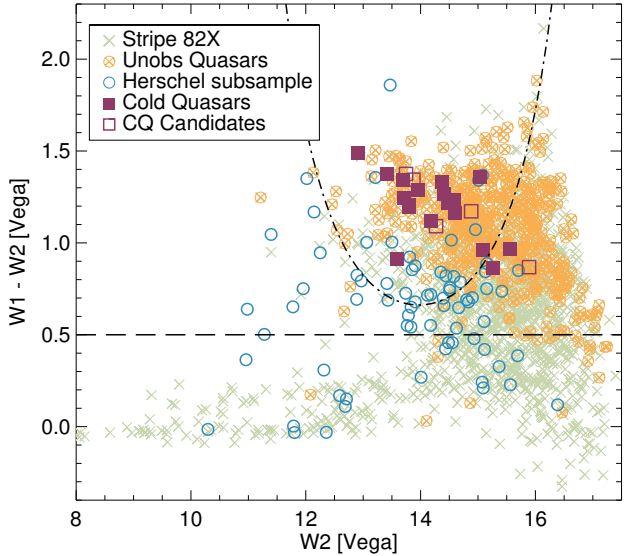}
\caption{Our Stripe 82X sample in WISE colorspace. Our {\it Herschel} subsample (blue circles) is a mix of galaxies dominated by AGN in the mid-IR as opposed to star formation. This is expected given the range of X-ray luminosities in this sample. All the Cold Quasars (solid purple squares) and Cold Quasar candidates (unfilled purple squares) lie above dashed line, which is defined as having a 90\% AGN selection completeness in \citep{assef2018}. Most meet the more stringent 90\% reliability selection (dot-dashed line) as well. From our parent Stripe 82X sample (green crosses), we plot unobscured quasars ($L_X>10^{44}\,$erg/s, $M_B<-23$) as the orange circles with crosses. The Cold Quasars are generally brighter and redder than the unobscured quasars. \label{wise}}
\end{figure}

Mid-IR color-color diagrams are an effective way of identifying luminous AGN \citep{lacy2004,stern2005,donley2012,mateos2012,stern2012}. The AGN typically has a torus that typically produces redder mid-IR colors than purely star-forming galaxies \citep{donley2012}. 
Twenty Cold Quasars are detected in the $W1$ and $W2$ bands, and all of these meet the WISE AGN selection criteria $W1-W2 > 0.5$ from \citet{assef2018}, which was designed to select AGN with a completeness of 90\% (dashed line; Figure \ref{wise}). They also satisfy the criteria
\begin{equation}
W1-W2 > 0.53\exp[0.183(W2-13.76)^2]
\end{equation}
from \citet{assef2018}, which was designed to select AGN with a reliability of 90\% (dot-dashed line; Figure \ref{wise}). In other words, our Cold Quasars have the near- and mid-IR emission expected of luminous AGN. A subset of mid-IR AGN are defined as hot dust-obscured galaxies \citep[Hot DOGS;][]{eisenhardt2012}. These are rare, dusty galaxies with $L_{\rm IR}>10^{13}\,L_\odot$ \citep{wu2012}. Hot DOGs are luminous quasars, yet dust obscured. Hot DOGs are also referred to as $W1W2$ dropouts due to their extremely red spectrum, and are typically selected according to $W1 > 17.4$ and $W2-W4 > 8.2$ \citep{eisenhardt2012}. None of our Cold Quasars meet the WISE selection criteria to be classified as Hot DOGs.

Of the parent unobscured quasar sample, 559 galaxies have $W1$ and $W2$ detections. Almost all meet the \citet{assef2018} completeness criterion shown in Figure \ref{wise}, but only 57\% satisfy the reliability criterion. 
Our Cold Quasars lie to the upper left in the distribution of unobscured quasars in Figure \ref{wise}. This means that they are in general both brighter in the near-IR ($W1$ and $W2$ trace near-IR emission at the redshifts of our sources) and redder than the average unobscured quasar in our survey. The median (standard deviation) $W1-W2$ color of the Cold Quasars is 1.22 (0.17), while the median (standard deviation) of the unobscured quasars is 1.11 (0.27).

\begin{figure}
\includegraphics[width=3.3in]{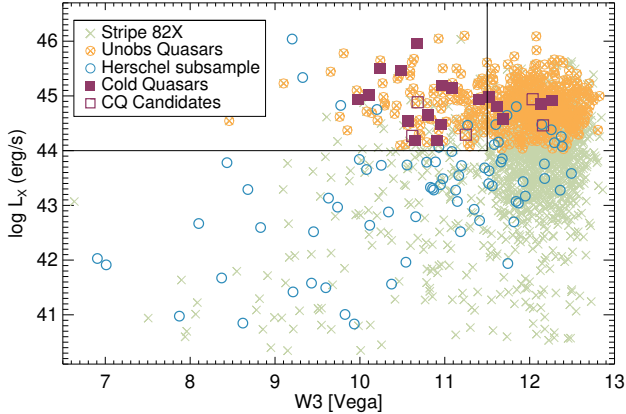}
\caption{Cold Quasars (purple squares) are brighter in the longer wavelength WISE bands than typical unobscured quasars (orange circles with crosses). The full {\it Herschel} subsample (blue circles) has a range of $W3$ magnitudes, but many of the lower $L_X$ sources are star-forming galaxies, which have enhanced $W3$ due to PAH emission. We recommend using $L_X > 1\times 10^{44}$\,erg/s and $W3 < 11.5$ (solid lines) to select Cold Quasar candidates.\label{selection}}
\end{figure}

The Cold Quasars are also more luminous in the longer wavelength WISE bands than the unobscured quasars (Figure \ref{selection}). At $z\sim1-2$, where the majority of our Cold Quasars lie, $W3$ spans $\lambda \sim4-6\,\mu$m, where the torus should outshine stellar emission. Through examining our sources, we find 13 Cold Quasars (out of 18 detected in WISE) satisfy the criterion
$W3< 11.5\, [{\rm Vega}] $. We determined this threshold by calculating the expected $W3$ emission of infrared-selected AGN at $z=3.2$, the redshift of the most distant Cold Quasar. We scale the Featureless AGN template from \citet{kirkpatrick2012}, which is empirically derived from IR AGN at $z=1-3$, to $S_{\rm 250\,\mu m}=30\,$mJy, the detection limit of the HerS survey, and calculate the observed $W3$ magnitude to be 11.5.

In contrast, only 19\% of the unobscured quasars have $W3< 11.5$.
Combining the X-ray and optical quasar criteria with the mid-IR criterion weeds out star-forming galaxies, which can be bright in W3 at $z\sim0-1$ due to the 6.2, 7.7, and 11.2\,$\mu$m PAH features falling in this bandpass. In fact, many of our {\it Herschel} subsample show significantly enhanced W3 emission, but their low $L_X$ ($<10^{42}\,$erg/s) and IR SED fitting indicate that they are star-forming galaxies. Due to the selection efficiency, the criteria
\begin{align}
L_X &>10^{44}\,{\rm erg/s} \\
M_B &<-23 \\
W3 &<11.5
\end{align} 
can be used to select Cold Quasar candidates for longer wavelength follow-up in future surveys.

\begin{figure}
\includegraphics[width=3.3in]{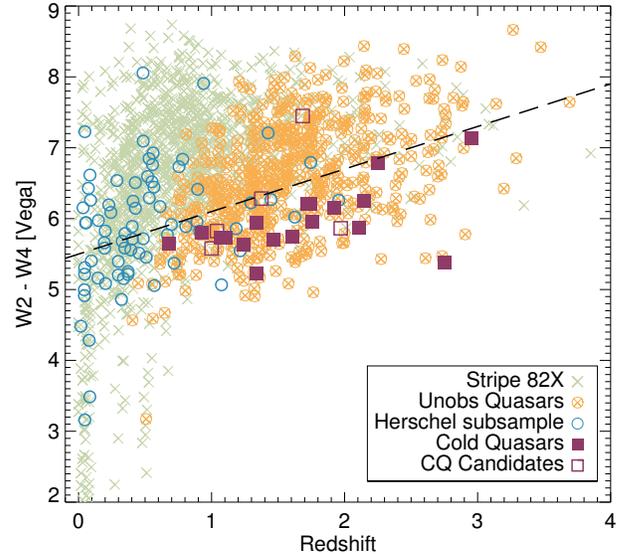}
\caption{We show the slope of the mid-IR emission, parameterized as $W2-W4$, as a function of redshift. Our Cold Quasars (purple squares) have bluer colors than the unobscured quasars in the parent sample (orange circles with crosses). All Cold Quasars lie below $W2-W4 = 0.6z +5.5$, plotted as the dashed line. The bluer colors may be due to a torus with a small optical depth, patchy circumnuclear dust, or increasing contribution to the mid-IR colors from star-forming regions. \label{slope}}
\end{figure}

The shape of the mid-IR SED is directly related to physical properties of the torus. In a two-component model, where the torus is composed of a smooth disk surrounding by clumps of dust, the slope of the mid-IR emission depends strongly on the optical depth of the both the disk and the clumps \citep{siebenmorgen2015}. We show the slope of the mid-IR SED, parameterized by the color $W2-W4$ as a function of redshift in Figure \ref{slope}. At $z\sim1-2$, $W4$ is tracing $\lambda\sim7-11\,\mu$m and $W2$ is tracing $\lambda\sim1.5-2.5\,\mu$m. Cold Quasars have a bluer $W2-W4$ color than the majority of unobscured quasars at similar redshifts. In fact, all Cold Quasars lie below the relation
\begin{equation}
W2-W4 = 0.6z +5.5
\label{w_eq}
\end{equation}
plotted as the dashed line. Only 40\% of the unobscured quasars lie below the line. This may indicate that Cold Quasars have different torus properties, specifically with a lower optical depth, than typical unobscured quasars, or a more face-on viewing angle. Alternately, if the primary obscuring material in blue quasars is circumnuclear, rather than a torus, this indicates a low covering fraction of circumnuclear dust. Of course, sources with a lot of star formation will have bluer $W2-W4$ colors at $z\sim1-2$, as $W4$ will be tracing the continuum in between the 11.2\,$\mu$m and 7.7\,$\mu$m PAH peaks, while $W2$ is tracing the stellar emission, which is a maximum around 1.6\,$\mu$m \citep{kirkpatrick2012}. Cold Quasars have high star formation rates and may be expected to have a significant contribution to their mid-IR emission from star formation, although this is not indicated by the SED fits. { More coverage of the mid-IR (e.g., with the {\it James Webb Space Telescope}) can tightly constrain how much is due to PAH emission.} 

Mid-IR bright AGN in star-forming host galaxies will have significant far-IR emission attributable to the host. \citet{kirkpatrick2013} examined the IR colors of dusty star-forming galaxies, some of which host AGN, from $z\sim1-3$. AGN are easily distinguished from star-forming galaxies using dust temperature, measured from the ratio of mid- to far-IR data. AGN-heated dust is much warmer than cool interstellar dust where stars form. With {\it Spitzer} and {\it Herschel}, the colors $S_{250{\rm \mu m}}/S_{24{\rm \mu m}}$ vs. $S_{8{\rm \mu m}}/S_{3.6{\rm \mu m}}$ reliably separate type-1 and type-2 AGN from star-forming galaxies by tracing the heating sources of the ISM and looking for hot dust from the torus, which produces a steeper continuum in AGN \citep{kirkpatrick2013,kirkpatrick2015}. 

\begin{figure}
\includegraphics[width=3.3in]{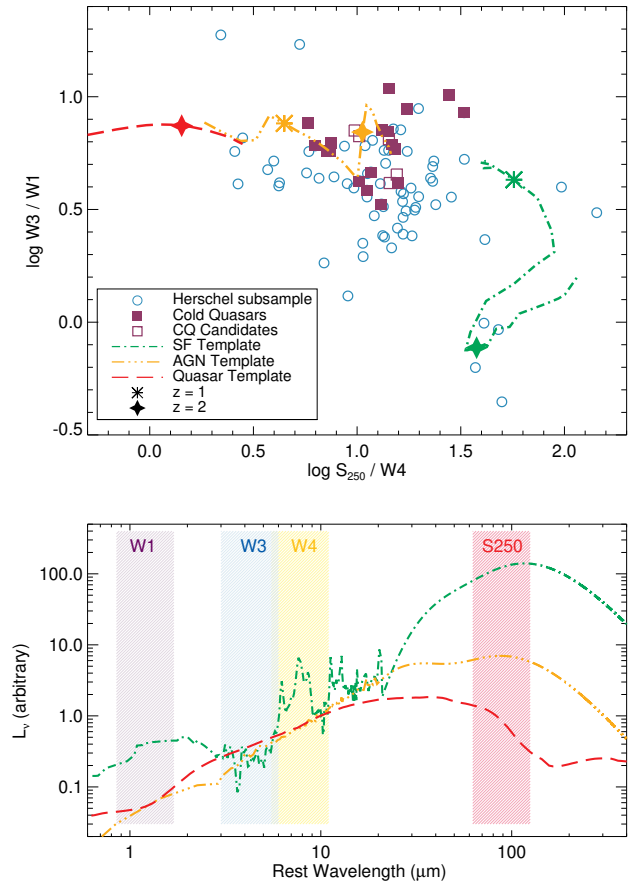}
\caption{{\it Top panel--}$S_{250}/W4$ traces the cold to warm dust, or average dust temperature, while $W3/W1$ traces the slope of the continuum emission. We plot redshift tracks from $z=0.5-3$ of a star-forming template (green dot-dashed line), AGN template \citep[orange dot-dash-dotted line][]{kirkpatrick2012}, and unobscured quasar template \citep[red dashed line][]{elvis1994} to illustrate how star-forming galaxies separate from AGN in this color diagram. Cold Quasars (purple filled squares) have similar dust continua but warmer dust than typical unobscured quasars, as do other Herschel-detected sources (blue circles). {\it Bottom panel--} We visually compare the three templates plotted in the top panel, normalized at 10\,$\mu$m. We illustrate with the shaded regions which portion of the SED is covered by the $W1$, $W3$, $W4$, and $S_{250}$ filters at $z=1-3$. \label{IR_color}}
\end{figure}

In place of {\it Spitzer}/MIPS and IRAC observations, we use $S_{250}/W4$ to trace the ratio of cold dust to warm dust, and $W3/W1$ to trace the slope of the mid-IR continuum, and we have converted all WISE Vega magnitudes to fluxes.
In the top panel of Figure \ref{IR_color}, we show where the Cold Quasars and {\it Herschel} subsample lie in this colorspace. We plot the redshift tracks from $z=0.5-3.0$ of the $z\sim1$ Star-Forming template and Featureless AGN template from \citet{kirkpatrick2012}. These templates were empirically created from 151 star-forming galaxies and AGN with {\it Spitzer} and {\it Herschel} observations at $z\sim1-3$. A few sources in our {\it Herschel} subsample lie in the region occupied by the star-forming template, and unsurprisingly, these are the sources with $L_X < 10^{42}$\,erg/s. The Cold Quasars all occupy the same region as the AGN template, while many of the {\it Herschel} subsample lie between the two regimes, indicated a mix of star formation and AGN emission in the mid-IR. The AGN template was derived from sources from the {\it Chandra} Deep Field South at $z\sim1-2$. These sources were identified as AGN on the basis of strong power-law emission in their {\it Spitzer}/IRS spectra. 75\% have $L_{\rm IR}>10^{12}\,L_\odot$. Of these 75\%, only 50\% have $L_X > 10^{44}$\,erg/s \citep{brightman2014}, and none meet the optical quasar definition, based on ACS photometry from the {\it Hubble Space Telescope} \citep{giavalisco2004} They are therefore a fundamentally different population than our Cold Quasars, and yet have similar ratios of far-IR emission.

We also plot a standard type-1 quasar template from \citet{elvis1994}. This template was created using the IRAS emission of UV and X-ray selected quasars, many of which are part of the 
Palomar-Green Survey \citep{green1986}. \citet{lyu2017} compare multiple type-1 quasar templates from the literature and determine that the templates of \citet{elvis1994}, 
\citet{symeonidis2016} and \citet{netzer2007}, which are all based on the Palomar-Green Survey, accurately represent the emission of unobscured quasars, while the \citet{kirkpatrick2012} 
templates are not valid for luminous type-1 objects. The Palomar-Green Survey is comprised mainly of low redshift quasars ($z<0.2$) which show a decline in far-IR emission beyond $
\lambda>20\,\mu$m. However, the \citet{elvis1994} template falls to the left of our sample, indicating that on average, low redshift unobscured quasars have much less cold dust than our Cold 
Quasars, which are all at $z>0.5$. In the bottom panel of Figure \ref{IR_color}, we compare the \citep{elvis1994} template with the AGN and star-forming template from \citet{kirkpatrick2012}. We illustrate with the shaded regions which portion of the SED is covered by the $W1$, $W3$, $W4$, and $S_{250}$ filters at $z=1-3$. 

Our Cold Quasars demonstrate that luminous, type-1 AGN can have an abundance of cold dust, albeit only rarely. The amount of cold dust may partially be 
attributed to the rising gas fractions of galaxies with redshift, so that local quasar templates cannot be compared to higher redshift quasars \citep[e.g.,][]{geach2011,kirkpatrick2017}.

\subsection{Are Cold Quasars Special?}
We have defined Cold Quasars as unobscured type-1 quasars with a substantial amount of cold dust ($S_{250}>30\,$mJy). Such galaxies have been observed before in the literature \citep{frayer1998,barvainis2002,omont2003,page2004,willott2007,coppin2008,stacey2010,simpson2019,ivison2019} so herein we discuss whether dusty, unobscured quasars deserve their own unique classification.
Our Cold Quasars all have intense SFRs, with the most extreme being $\sim400\,M_\odot$/yr. But, does it make sense to identify these quasars as a separate class? Answering that requires that we understand the expected amount of infrared emission in unobscured quasars, in itself a difficult question. There are 805 Stripe 82X sources with $L_X>10^{44}\,$erg/s and $M_B <-23$. Cold Quasars are 3\% of this population. If we include the candidates, this rises to 4\%. Based on detectability in the far-IR, Cold Quasars are indeed a special, rare population. Of course, these numbers depend sensitively on the depth of the {\it Herschel} data in Stripe 82, and a deeper survey may uncover more Cold Quasars. 

The relatively low sensitivity of {\it Herschel} means that only the most luminous infrared galaxies are detected with increasing redshift, even in the deepest fields \citep[e.g.][]{elbaz2011}. One way around {\it Herschel} detection limits is through stacking. \citet{stanley2017} compile composite IR SEDs by stacking {\it Herschel} images and averaging {\it WISE} fluxes for 3000 optical SDSS unobscured AGN. { They measure the average AGN bolometric luminosity, $L_{\rm Bol}$, using the optical spectroscopic values from the SDSS DR7 catalog in \citet{shen2011} and the stellar mass using $M_\bullet$ from \citet{shen2011} and Equation \ref{eqbol}. \citet{stanley2017} determine the SFR by decomposing the stacked WISE and {\it Herschel} photometry using the {\tt DecompIR} code from \citet{mullaney2011}. As discussed in \citet{kirkpatrick2012} and \citet{kirkpatrick2015}, the Featureless AGN template we use to decompose the Cold Quasars is remarkably similar to the \citet{mullaney2011} AGN model. 

The \citet{stanley2017} sample is unbiased with respect to IR detections, and so we presume that their SFRs represent the typical SFR of unobscured quasars of a given bolometric luminosity. 
We compare our Cold Quasars and {\it Herschel} subsample with the \citet{stanley2017} sample in Figure \ref{lir_lx}. We find that our Cold Quasars have a 50\% higher SFR 
for a given $L_{\rm Bol}$. To make this more obvious, we plot the ratio of SFR to black hole accretion rate (BHAR) as a function of $L_{\rm Bol}$. We calculate BHAR as
\begin{equation}
{\rm BHAR} = \frac{(1-\eta)L_{\rm bol}}{\eta c^2}
\end{equation}
where $\eta=0.1$ is the radiative efficiency \citep{soltan1982}. It is important to bear in mind that the stars in Figure \ref{lir_lx} are averages of $\sim90$ quasars each, where 
the error bars represent the standard deviation. Our Cold Quasars lie above these averages and standard deviations.

\begin{figure}
\includegraphics[width=\linewidth]{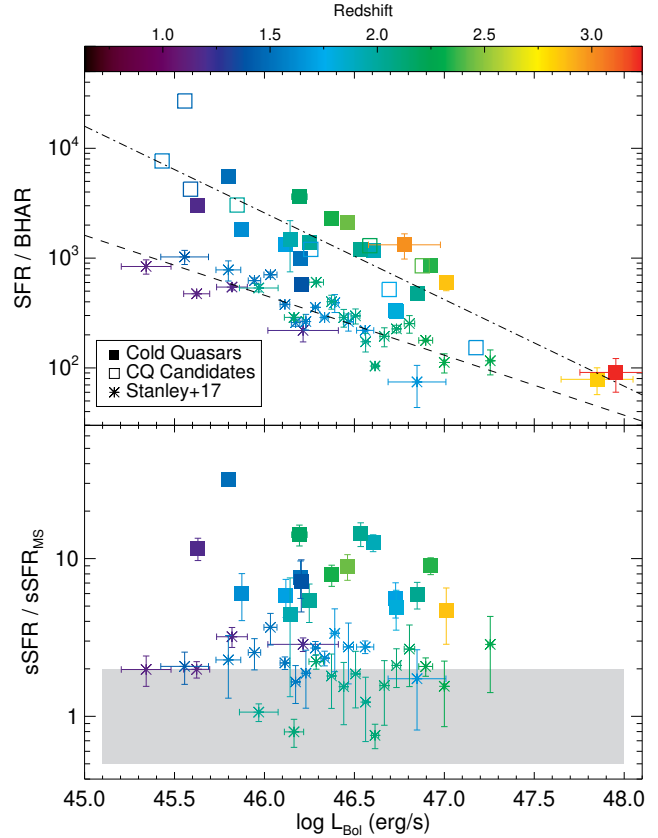}
\caption{{\it Top}--$L_{\rm Bol}$ vs. SFR/BHAR for the Cold Quasars (filled squares), Cold Quasar candidates (unfilled squares), and stacked unobscured quasars from \citet{stanley2017}. Colors correspond to redshift. The stars are the averages of $\sim90$ quasars each. On average, the Cold Quasars have 50\% as much star formation as the unobscured quasars at the same bolometric luminosities. The dashed line shows the fit to the stars, while the dot-dashed line is the fit to the Cold Quasars and Cold Quasar candidates.{\it Bottom}--$L_{\rm Bol}$ vs. SFR/SFR$_{\rm MS}$, where SFR$_{\rm MS}$ is the SFR a galaxy on the main sequence would have, accounting for both $M_\ast$ and $z$. The shaded region shows a factor of two above and below the main sequence. There is no trend with $L_{\rm Bol}$. The Cold Quasars on average lie $9\times$ above the main sequence, while the \citet{stanley2017} unobscured quasars lie $\sim2\times$ above the main sequence. \label{lir_lx}}
\end{figure}

Galaxy SFR and evolution are commonly understood in the context of the main sequence, which is the tight relationship between SFR and $M_\ast$ exhibited by most galaxies \citep{elbaz2007,noeske2007}. In the major merger paradigm, mergers trigger starbursts, significantly elevating SFR above the main sequence \citep{sanders1996}. The quasar phase is predicted to follow the starburst phase \citep{hopkins2007}. For typical unobscured quasars, \citet{stanley2017}
find no significant difference between their SFRs and those of star-forming galaxies on the main sequence at similar redshifts. We calculate the distance of the Cold Quasars and the \citet{stanley2017} stacks from the main sequence. The main sequence evolves with redshift, and it flattens at higher $M_\ast$. We use the relationship between $M_\ast$ and SFR parameterized in \citet{lee2015} at $z=1.2$:
\begin{equation}
\log {\rm SFR_{MS}} = 1.72 - \log \left[1 + \left(\frac{M_\ast}{2\times10^{10} M_\odot}\right)^{-1.07}\right]
\end{equation}
and then we renormalize depending on redshift \citep{speagle2014}:
\begin{equation}
{\rm SFR_{MS}}(z) = \left(\frac{1+z}{1+1.2}\right)^{2.9}\times{\rm SFR_{MS}(z=1.2)}
\end{equation}
We plot ${\rm SFR}/{\rm SFR_{MS}}$ as a function of $L_{\rm Bol}$ in the bottom panel of Figure \ref{lir_lx}. The shaded region illustrates a factor of 2 above and below the main sequence. Dusty star-forming galaxies with $L_{\rm IR}=10^{11}-10^{12}\,L_\odot$ typically lie on the main sequence \citep{casey2014}. On average, our Cold Quasars have ${\rm SFR}/{\rm SFR_{MS}}=9.3$, with a standard deviation of 6.4. The \citet{stanley2017} sample have a weighted mean of ${\rm SFR}/{\rm SFR_{MS}}=2.1$, where the points are weighted by the number of galaxies comprising each one. The Cold Quasars are $9\times$ above the main sequence, well into the starburst regime \citep[e.g.][]{elbaz2011}. Cold Quasars can be understood to be unobscured, type-1 quasars that lie in the starbursting regime of the 
main sequence. Again, we remind the reader that our SFRs are likely lower limits, since the choice of AGN template includes a significant amount of far-IR heating. Additional observations with far-IR telescopes (e.g., NASA/SOFIA) can help constrain this. Our Cold Quasars have nearly 5$\times$ as much star formation than the average unobscured quasar at the same bolometric luminosity.
}

\begin{figure}
\includegraphics[width=3.3in]{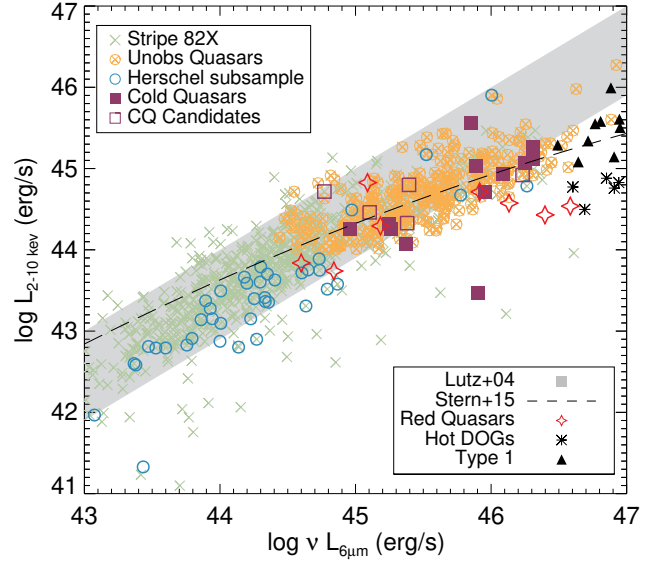}
\caption{Hard X-ray luminosity as a function of 6\,$\mu$m luminosity. We plot the relationship derived from local galaxies (grey shaded region), the relationship derived from higher luminosity, higher redshift sources (dashed line), and several different populations compiled in \citet{glikman2018}. The Cold Quasars predominantly follow the expected relations and have more X-ray emission relative to the Red Quasars and Hot DOGs. This indicates less gas absorption in the Cold Quasars. \label{x6}}
\end{figure}

Finally, we consider whether Cold Quasars show the expected amount of mid-IR emission given their X-ray luminosity in Figure \ref{x6}. We calculated $\nu L_{6\,\mu \rm m}$ by interpolating between the WISE fluxes. {We note that the AGN dominates the emission at these wavelengths (see fits in the Appendix), so there is no need to remove the host contribution before interpolating.} The relationship between mid-IR luminosity (parameterized by $\nu L_{\rm 6\mu m}$) and hard (2-10\,kev) X-ray luminosity has been quantified for local sources \citep{lutz2004} and high luminosity sources \citep{stern2015}. Sources that lie below the expected relationship (shaded region and dashed line) are under-luminous in the X-ray. They can be either Compton thick or perhaps accreting above the Eddington limit, which may be the case for Hot DOGs \citep{ricci2017a,wu2018}. However, except for one source, our Cold Quasars lie around the expected relations derived in both \citet{lutz2004} and \citet{stern2015}. The outlier is source 507, which has a soft X-ray luminosity (0.5-2 keV) four times greater than the hard X-ray luminosity, but a blue optical spectrum. It is possible that the hard X-ray luminosity has been underestimated in this source. The Cold Quasars have enhanced X-ray emission compared to the Red Quasars (green stars) and Hot DOGs (black asterisks) from \citet{lamassa2017b,glikman2017,glikman2018}, particularly at $\nu L_6 > 10^{46}$\,erg/s. However, they follow the same trend as type-1 quasars \citep[black triangles;][]{glikman2018}. $\nu L_{6}$ traces the amount of hot dust from the torus, while $L_X$ is sensitive to gas absorption. For a given $\nu L_{6}$, Cold Quasars have less gas absorption than Red Quasars. Similar to the Red Quasars, local merging ultra luminous galaxies are also underluminous in the X-rays compared with the infrared, indicating a high degree of absorption attributable to the merger \citep{teng2015}.

In the major merger framework, obscured quasars are understood to be a specific phase of a massive galaxy's life right before blowout \citep{hopkins2007,kocevski2015,glikman2015}. 
Tidal tails and train wreck merger signatures are still visible in Red Quasars, which have gas absorption visible through lower X-ray emission and redder optical continua \citep{glikman2015}. High gas column densities are also observed in local major mergers of massive dusty galaxies, many of which harbor obscured AGN \citep{petric2011,teng2015}. On the other hand, \citet{dai2014} identify a large population of blue broad-line quasars at $z\sim1-3$ detected at 24\,$\mu$m which have clear merger signatures, illustrating that 
mergers do not universally obscure the central AGN. 
Within the merging paradigm, the rare Cold Quasars we have defined are perhaps in a critical transition phase where the AGN growth co-exists peacefully within a starbursting galaxy for a brief epoch before quenching. 
Cold Quasars may immediately follow the Red Quasar stage, where a quasar wind has swept through the inner galaxy, removing the dust, but has not yet reached the star forming outskirts. 

Crucial to understanding the evolutionary stage of Cold Quasars is understanding where the dust lies and the morphology of the host galaxy. ALMA observations of submillimeter galaxies at $z\sim2-3$ have revealed that the dust is compact, more so than the gas, mainly lying within 1--3 kpc of the galaxy center \citep{hodge2016,rujopakarn2016,hodge2019,rujopakarn2019}. If the same is true for Cold Quasars, then perhaps the quasar has managed to punch a hole in the dust, producing a blue color. In that case, Cold Quasars may simply be patchy Red Quasars, rather than a separate evolutionary phase altogether. Red Quasars have clear merger signatures, so one way to test this hypothesis is to obtain sensitive observations of the ISM, where the quasar will not outshine the host galaxy, and look for disturbed morphological features. On the other hand, the quasar source may not be cospatial with the far-IR source responsible for the {\it Herschel/SPIRE} emission. If Cold Quasars are in the middle of a merger, the quasar portion could be significantly offset from the wet portion of the merger responsible for the burst of star formation \citep{fu2017}.
Currently, the spatial extent of submillimeter emission from quasars is unknown. ALMA observations of the spatial distribution of the ISM in these galaxies is required to both look for fueling signatures, such as tidal tails, and determine the location of the dust and gas.

If indeed Red Quasars, Cold Quasars, and Blue Quasars are linked together in an evolutionary sequence, the life cycle may be 3-4\% of the blue quasar phase, which is loosely constrained to be $10^6-10^9$\,yr \citep{conroy2013,laplante2016}. For comparison, the Red Quasar phase is 20\% of the blue quasar phase \citep{glikman2012}. We have based our estimation on considering only those unobscured quasars that lie in the same redshift range as our Cold Quasars. Tighter constraints can be placed with a mass-matched sample, but we currently lack stellar mass estimates for the full Stripe 82X catalog. If the blue quasar phase lasts a Gyr, then Cold Quasars may last only 40 Myr. 

\section{Conclusions}
\label{sec:conc}
We present a rare sample of X-ray and optically selected broad-line quasars that are detected in all three {\it Herschel}/SPIRE bands. We define Cold Quasars to be broad-line AGN with $L_X > 10^{44}\,$erg/s, $M_B < -23$, and $S_{250}>30\,$mJy. Our findings are:
\begin{enumerate}
\item Cold Quasars and Cold Quasar candidates comprise 4\% of the unobscured quasar population in Stripe 82X. If understood through the merger-driven paradigm, the duration may be 4\% of the blue quasar phase.
\item Cold Quasars have enhanced mid-IR emission relative to most unobscured quasars. $W3 < 11.5$\,[Vega] is an efficient mid-IR selection criteria for Cold Quasar candidates, selecting 72\% of our Cold Quasars but only 19\% of the full unobscured quasar population. This implies they have considerably more dust and possibly less circumnuclear obscuration than typical unobscured quasars.
\item Cold Quasars also have a bluer mid-IR spectra than the average unobscured quasar, and the color $W4-W2$ can also be used to select Cold Quasar candidates. This blue color may be due to a torus with a lower optical depth than is typical of unobscured quasars, or it may be due to contamination of the mid-IR by the strong star formation emission in these galaxies.
\item Cold Quasars have significantly more cold dust, as traced by $S_{\rm 250\,\mu m}$, than expected based on lower redshift unobscured quasars. The amount of cold dust is consistent with that measured in mid-IR selected AGN at $z\sim1-2$. The cold dust can be attributed to star formation, and Cold Quasars have high SFRs of $\sim100-400\,M_\odot$/yr.
\item The substantial SFRs of Cold Quasars are nearly 5$\times$ higher than the average unobscured quasar at the same redshift. Cold Quasars lie well off the main sequence (on average, their SFRs are 9$\times$ higher than the main sequence) and qualify as starburst galaxies. Whether these immense AGN luminosities and SFRs are triggered by major mergers will require high resolution submm follow-up with ALMA.
\item Cold Quasars are best defined as blue type-1 quasars that exist in starbursting galaxies.
\end{enumerate}

This exciting population of unobscured quasars with significant amounts of cold dust could represent an transition phase between dust-obscured quasars an unobscured quasars that have already depleted their interstellar medium. Further observations of the ISM will provide a more complete picture of how close these sources are to quenching.

\acknowledgements
We thank the anonymous referee for their helpful direction in improving this work. AK thanks Rob Ivison and Ian Smail for their generous insights and conversations which have added detail to this paper.
This research is based upon work supported by NASA under award No.
80NSSC18K0418 to Yale University and by the National Science Foundation under
Grant No. AST-1715512. E.T. acknowledges support from FONDECYT Regular 1160999 and 1190818, CONICYT PIA ACT172033 and Basal-CATA AFB170002 grants. A.K. gratefully acknowledges support from the YCAA Prize Postdoctoral Fellowship. E.G. acknowledges the generous support of the Cottrell College Award through the Research Corporation for Science Advancement.

\clearpage

\appendix
We show all the spectra in Figure \ref{spec1} from SDSS with the relevant broad lines indicated. We discuss spectral fitting and derived quantities in a future paper (Estrada et al. 2020, in prep).

\begin{figure*}[ht!]
\includegraphics[width=2.4in]{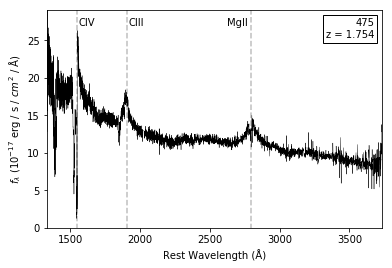}
\includegraphics[width=2.4in]{507-Full.png}
\includegraphics[width=2.4in]{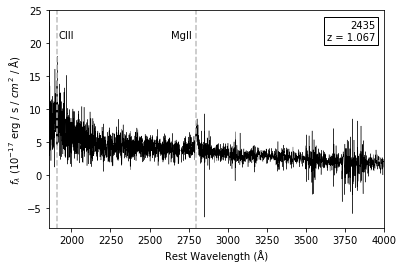}
\includegraphics[width=2.4in]{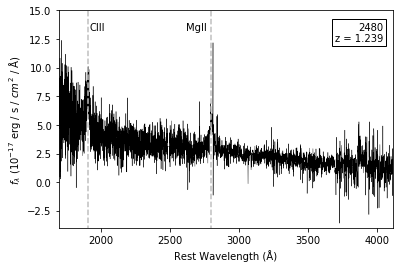}
\includegraphics[width=2.4in]{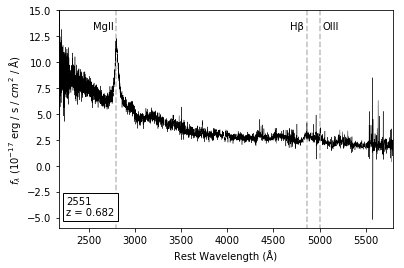}
\includegraphics[width=2.4in]{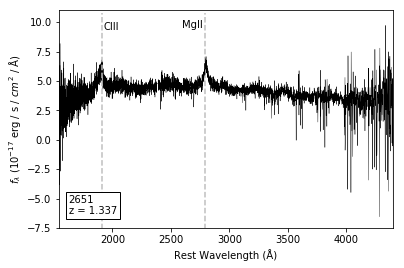}
\includegraphics[width=2.4in]{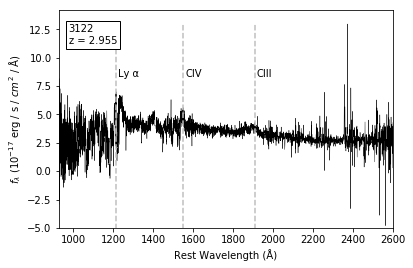}
\includegraphics[width=2.4in]{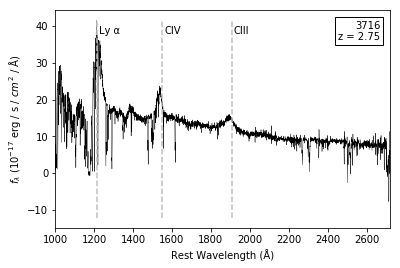}
\includegraphics[width=2.4in]{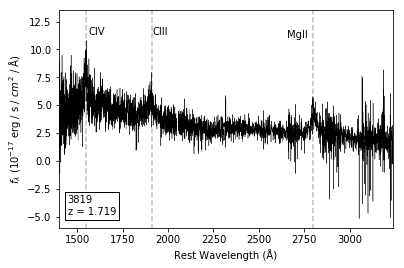}
\includegraphics[width=2.4in]{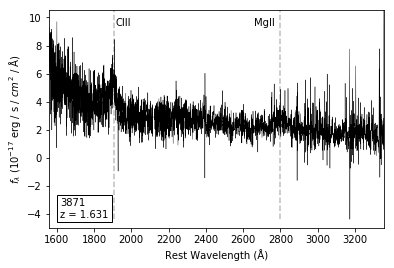}
\includegraphics[width=2.4in]{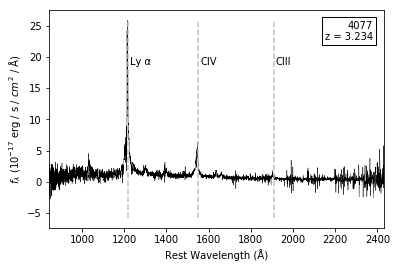}
\includegraphics[width=2.4in]{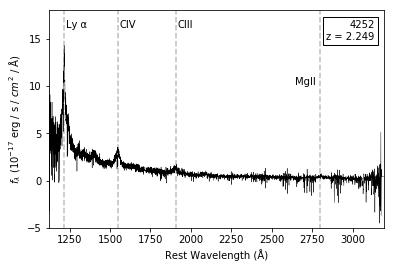}
\caption{SDSS spectra of the Cold Quasars with the relevant lines marked. \label{spec1}}
\end{figure*}

\setcounter{figure}{10} 
\begin{figure*}[ht!]
\includegraphics[width=2.4in]{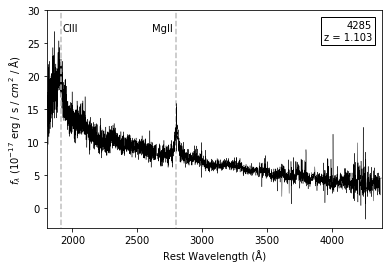}
\includegraphics[width=2.4in]{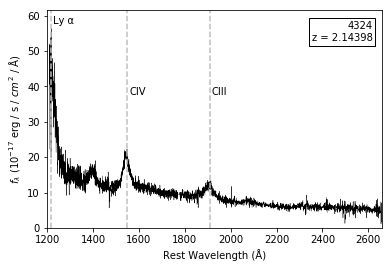}
\includegraphics[width=2.4in]{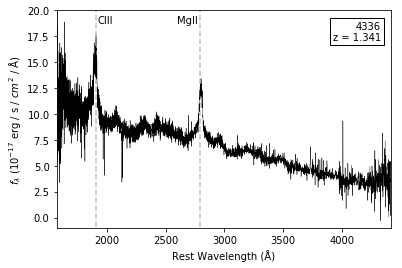}
\includegraphics[width=2.4in]{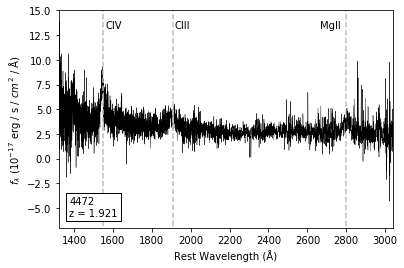}
\includegraphics[width=2.4in]{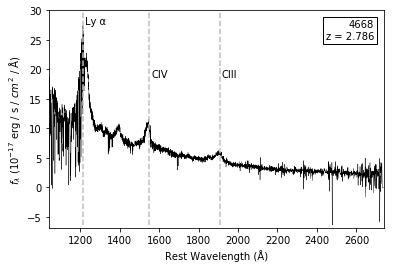}
\includegraphics[width=2.4in]{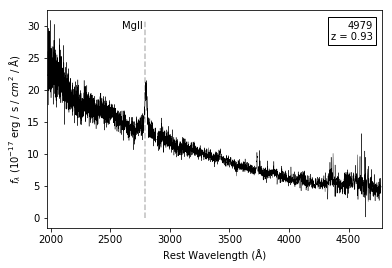}
\includegraphics[width=2.4in]{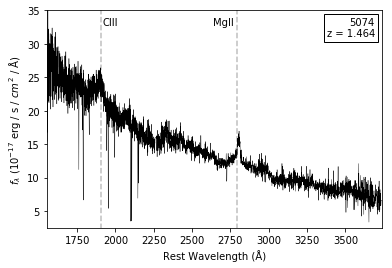}
\includegraphics[width=2.4in]{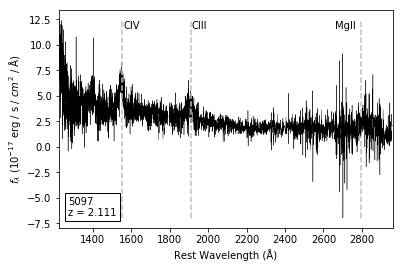}
\includegraphics[width=2.4in]{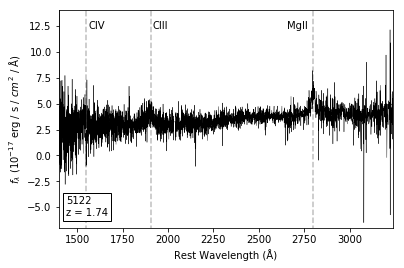}
\caption{{\it (continued)}}
\end{figure*}

\clearpage

Here we show the SED decomposition used to obtain $L_{\rm IR}$, SFR, and $f_{\rm AGN}$ in Figure \ref{sed3fit}. Each set of WISE and {\it Herschel} photometry is fit by a combination of the Featureless AGN template from \citet{kirkpatrick2012} and the submillimeter galaxy template from \citet{pope2008}. See details in Section \ref{fitmethod}.

\begin{figure*}[ht!]
\includegraphics[width=2.4in]{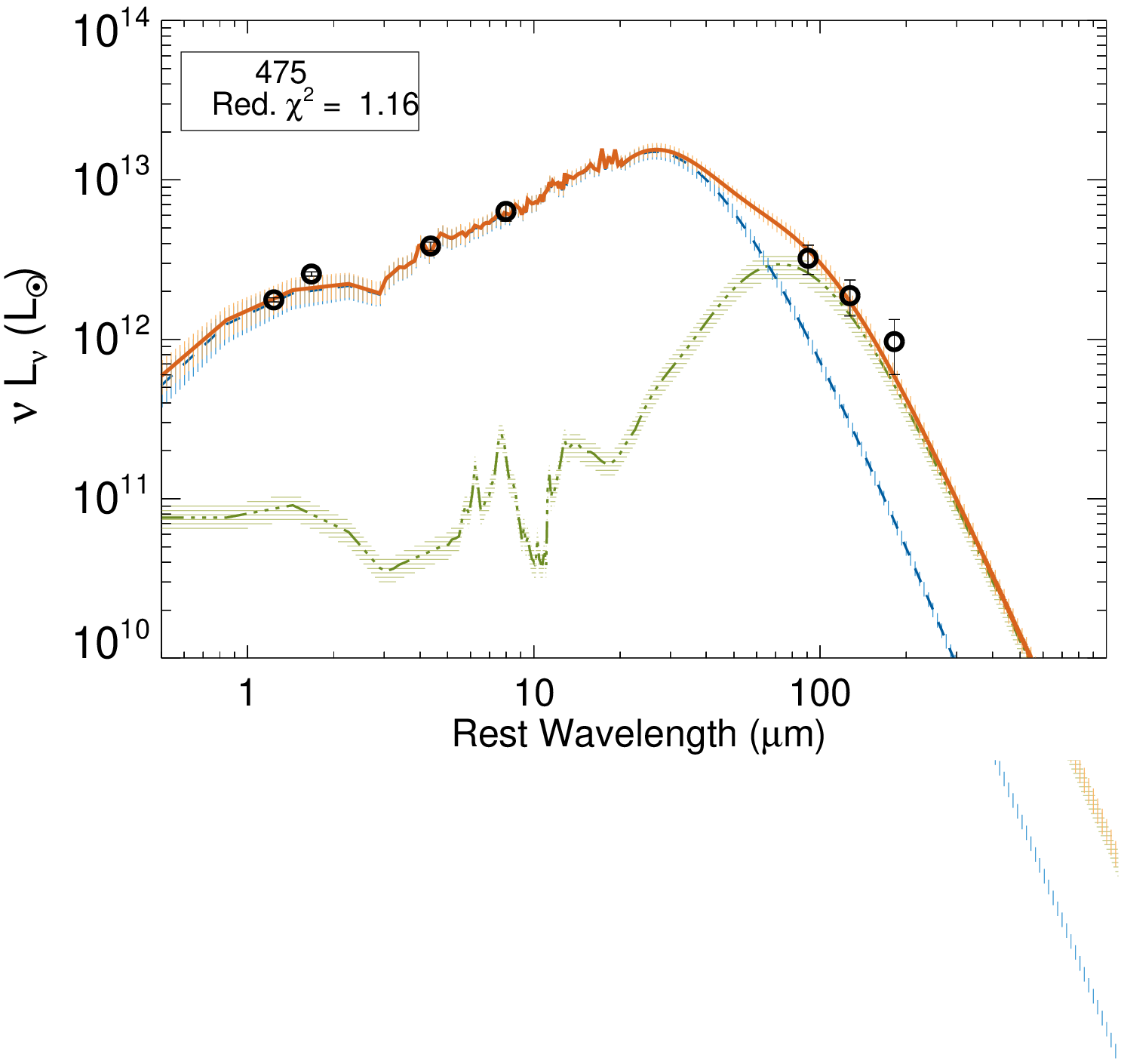}
\includegraphics[width=2.4in]{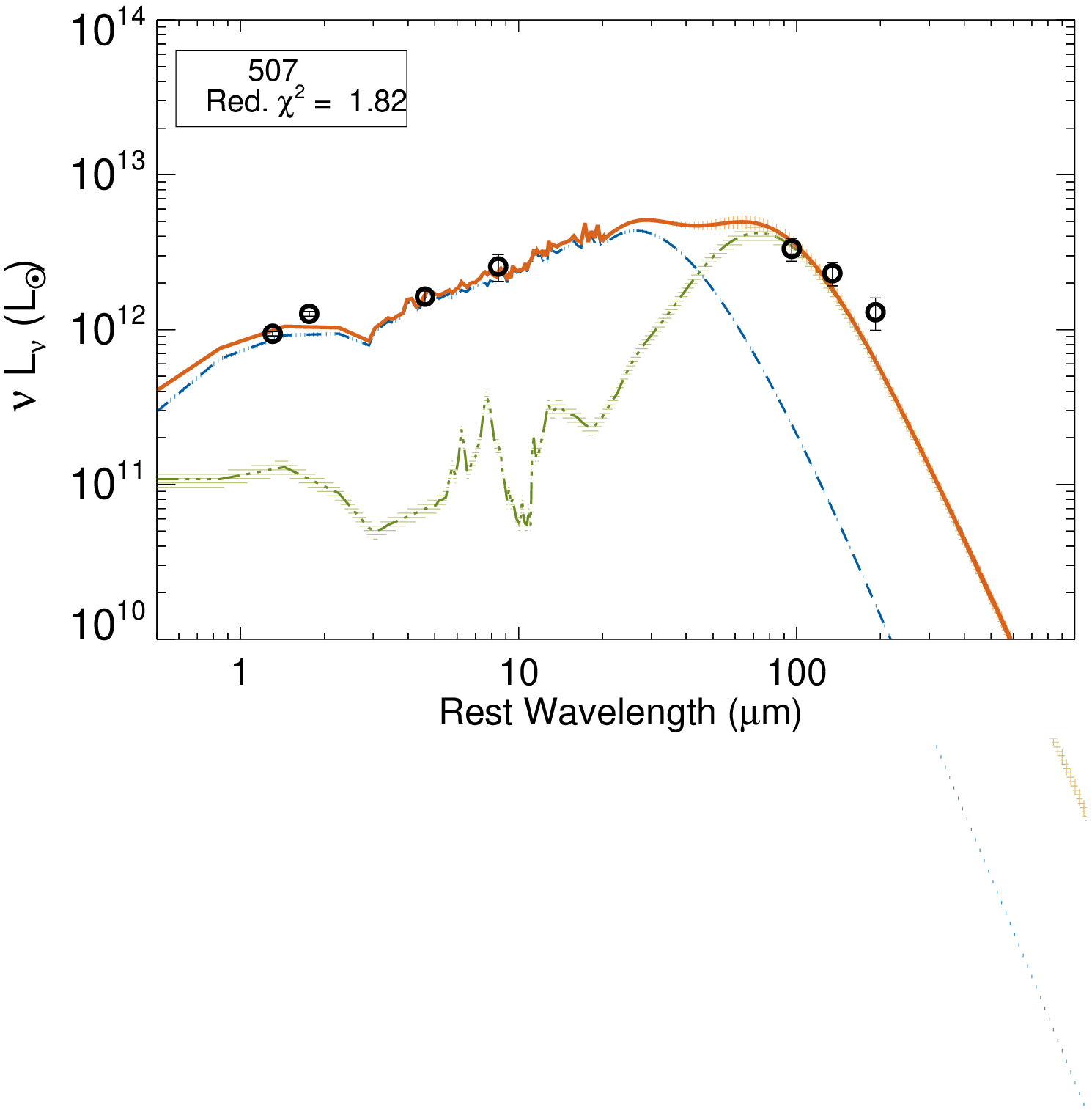}
\includegraphics[width=2.4in]{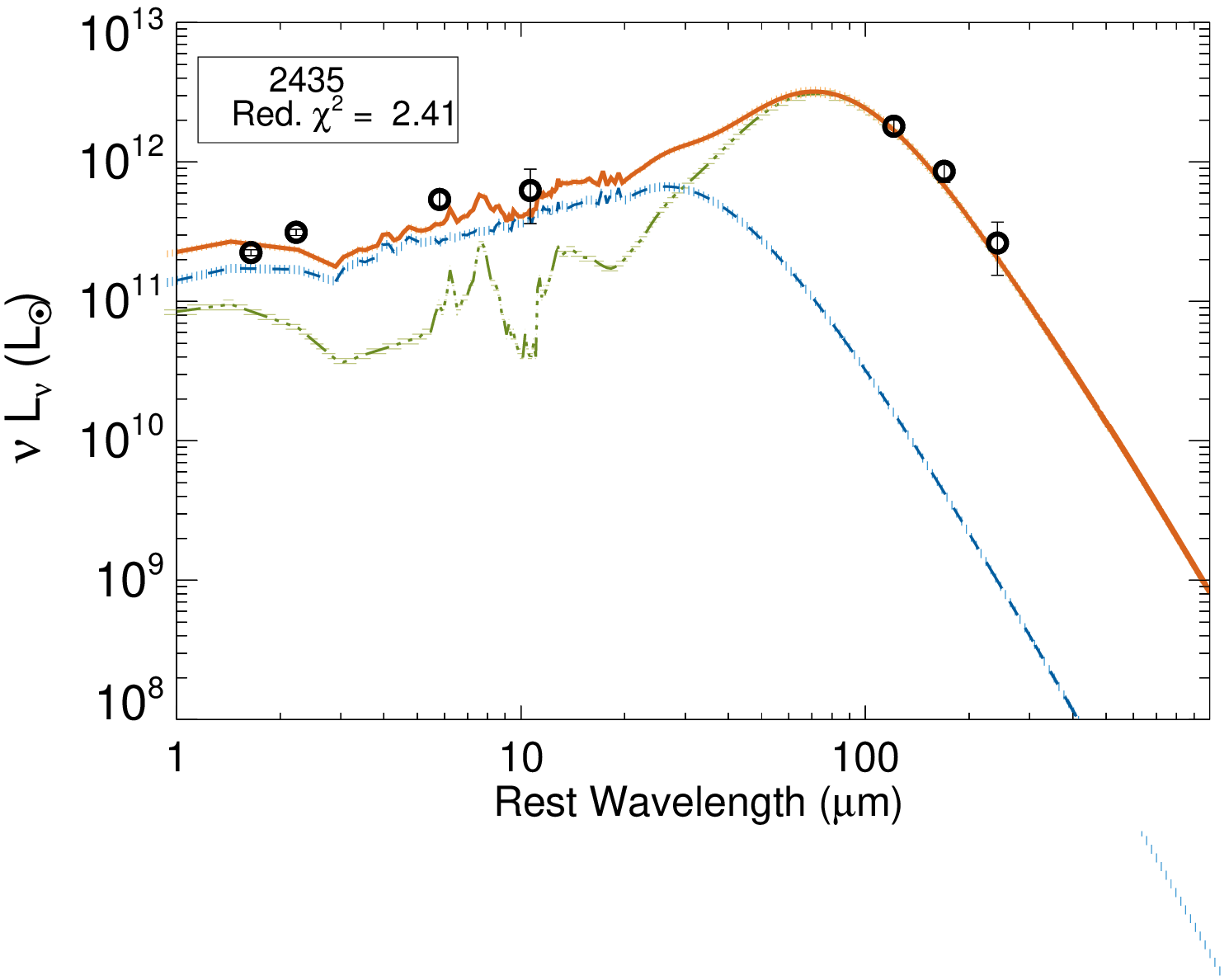}
\includegraphics[width=2.4in]{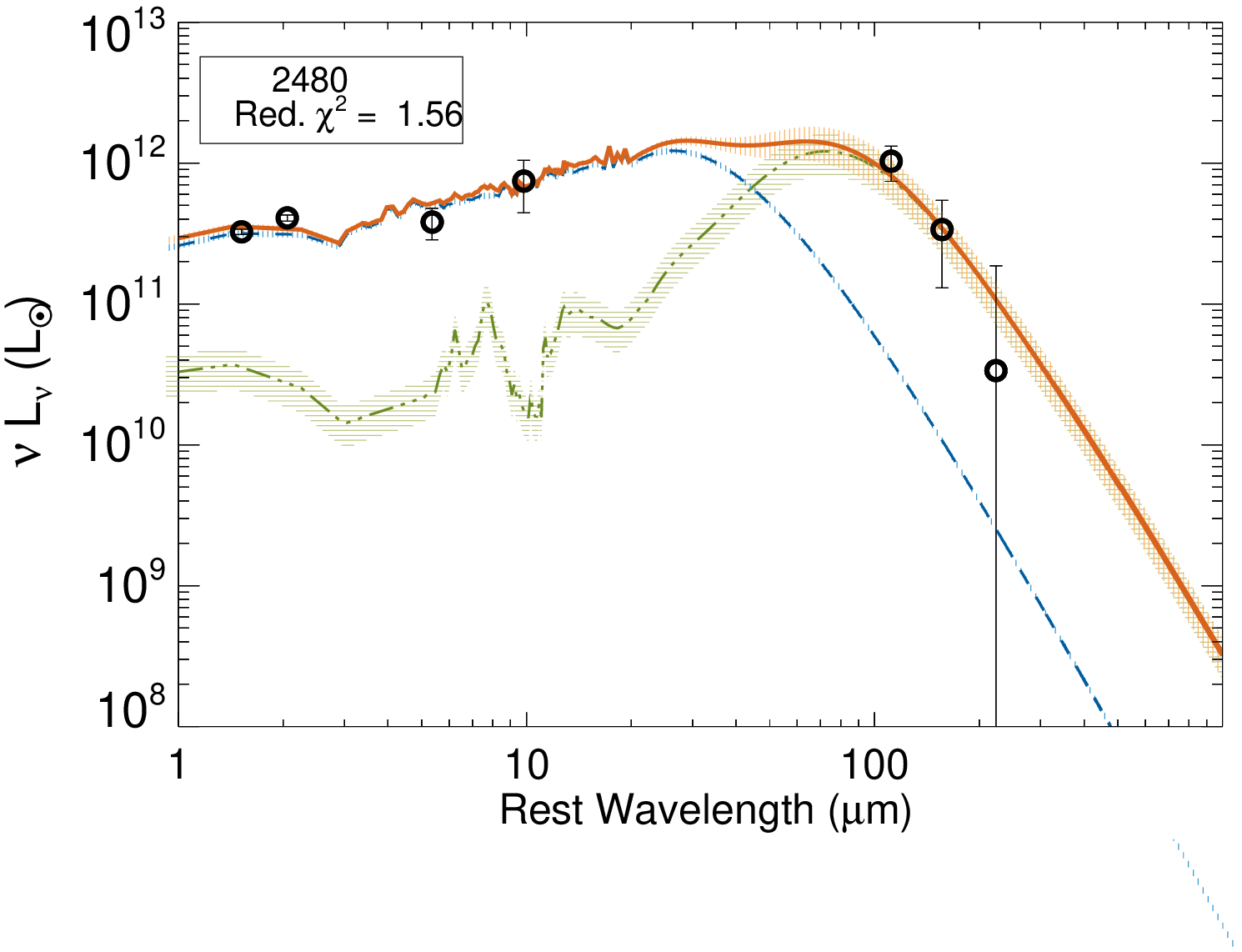}
\includegraphics[width=2.4in]{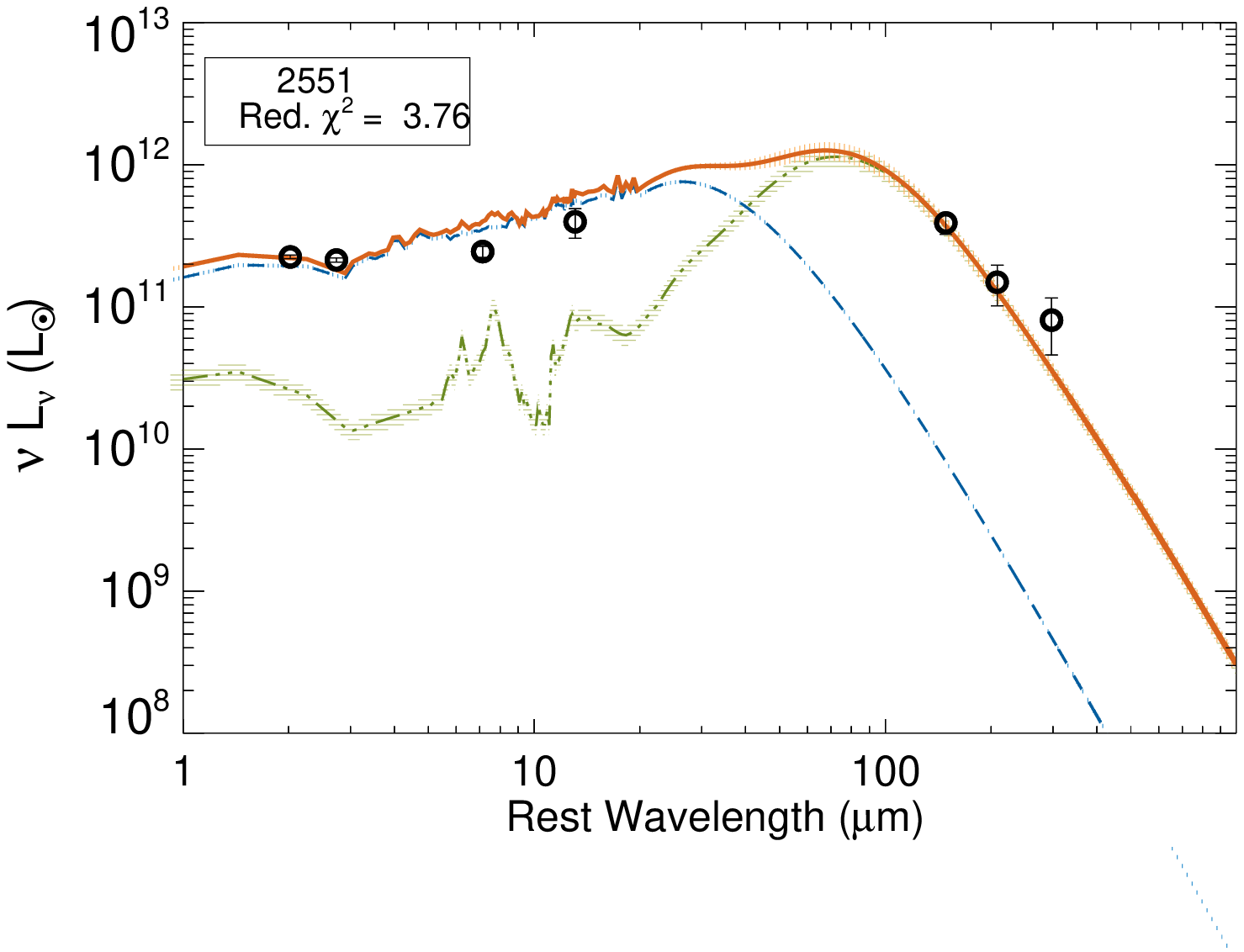}
\includegraphics[width=2.4in]{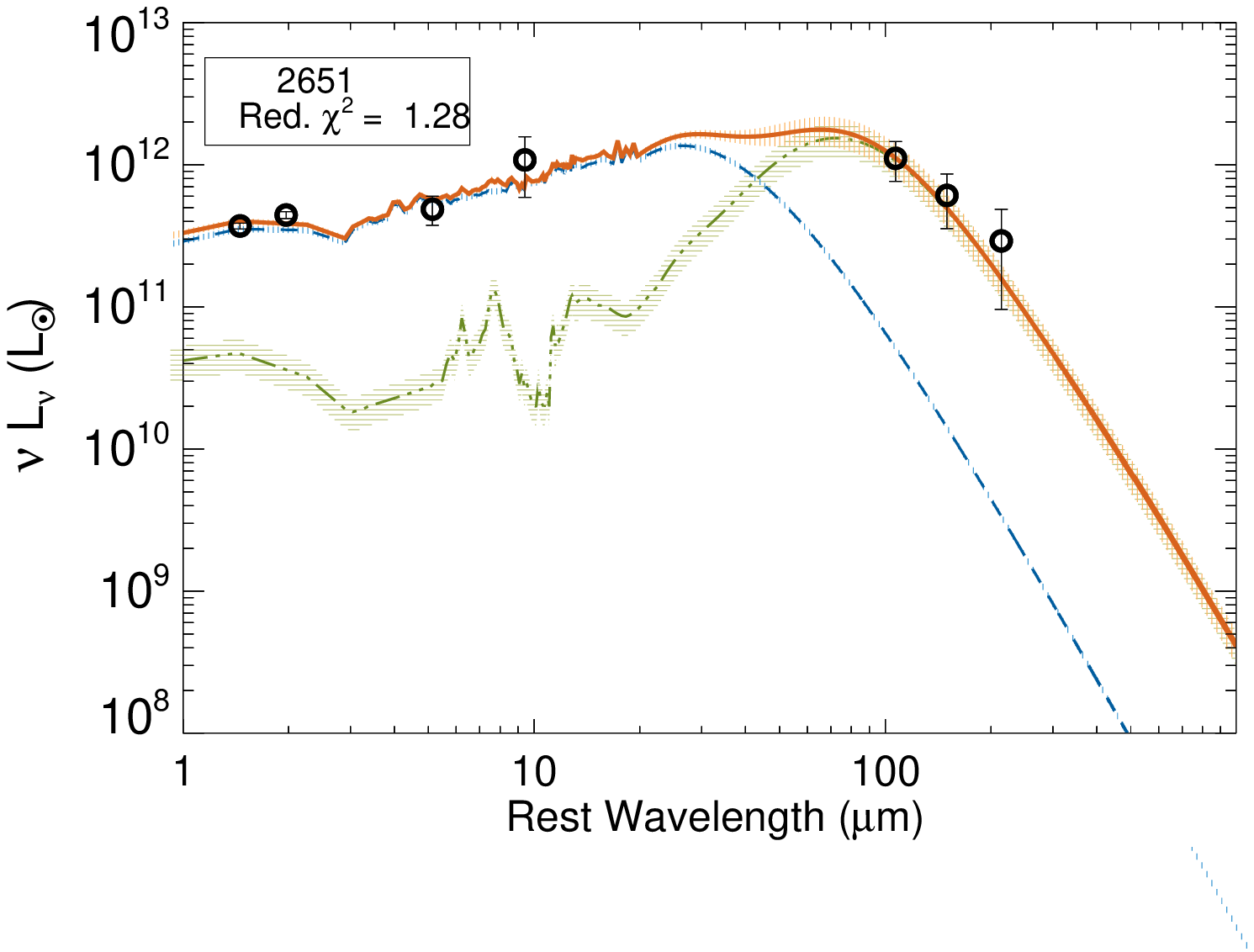}
\includegraphics[width=2.4in]{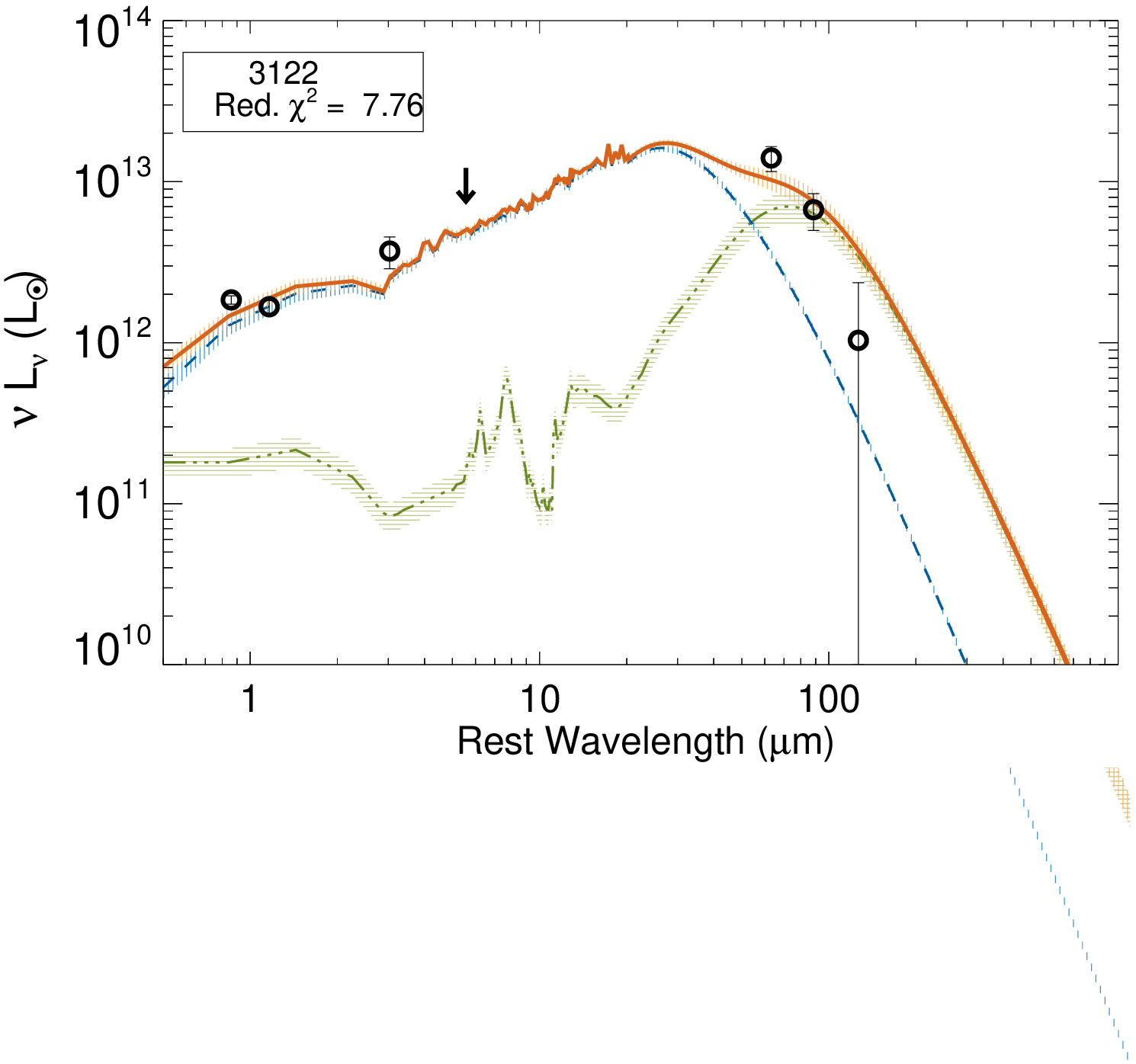}
\includegraphics[width=2.4in]{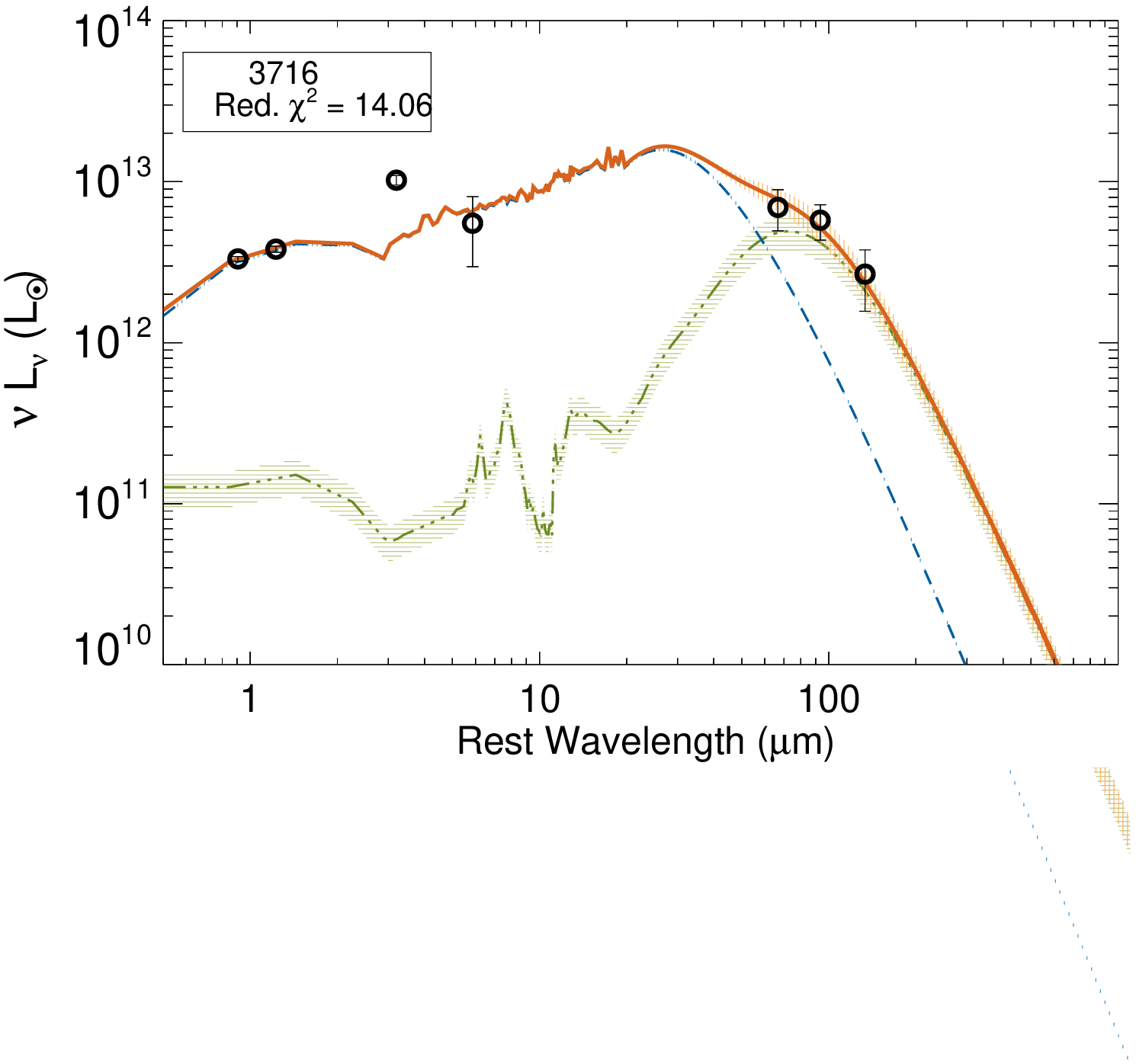}
\includegraphics[width=2.4in]{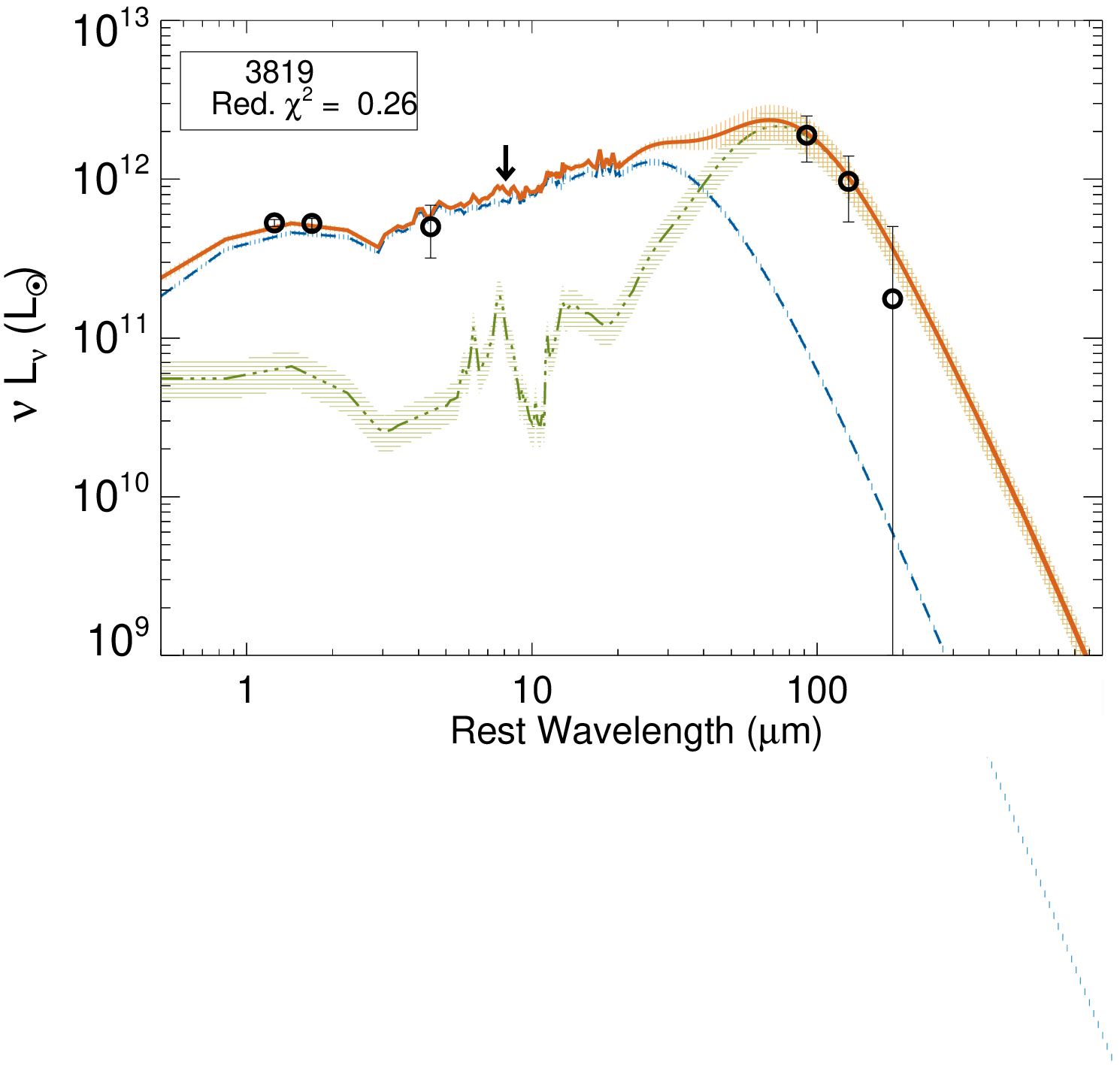}
\includegraphics[width=2.4in]{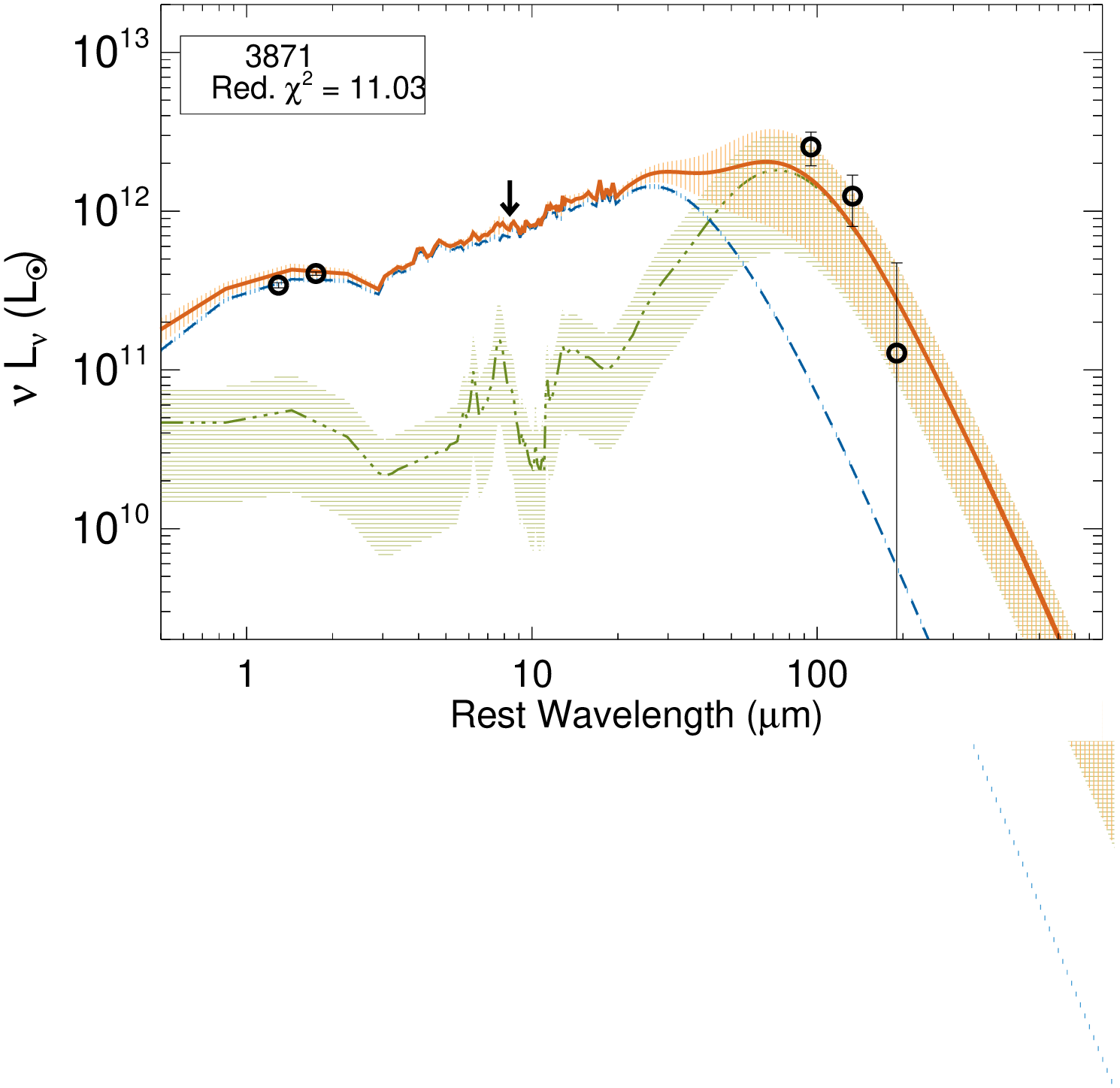}
\includegraphics[width=2.4in]{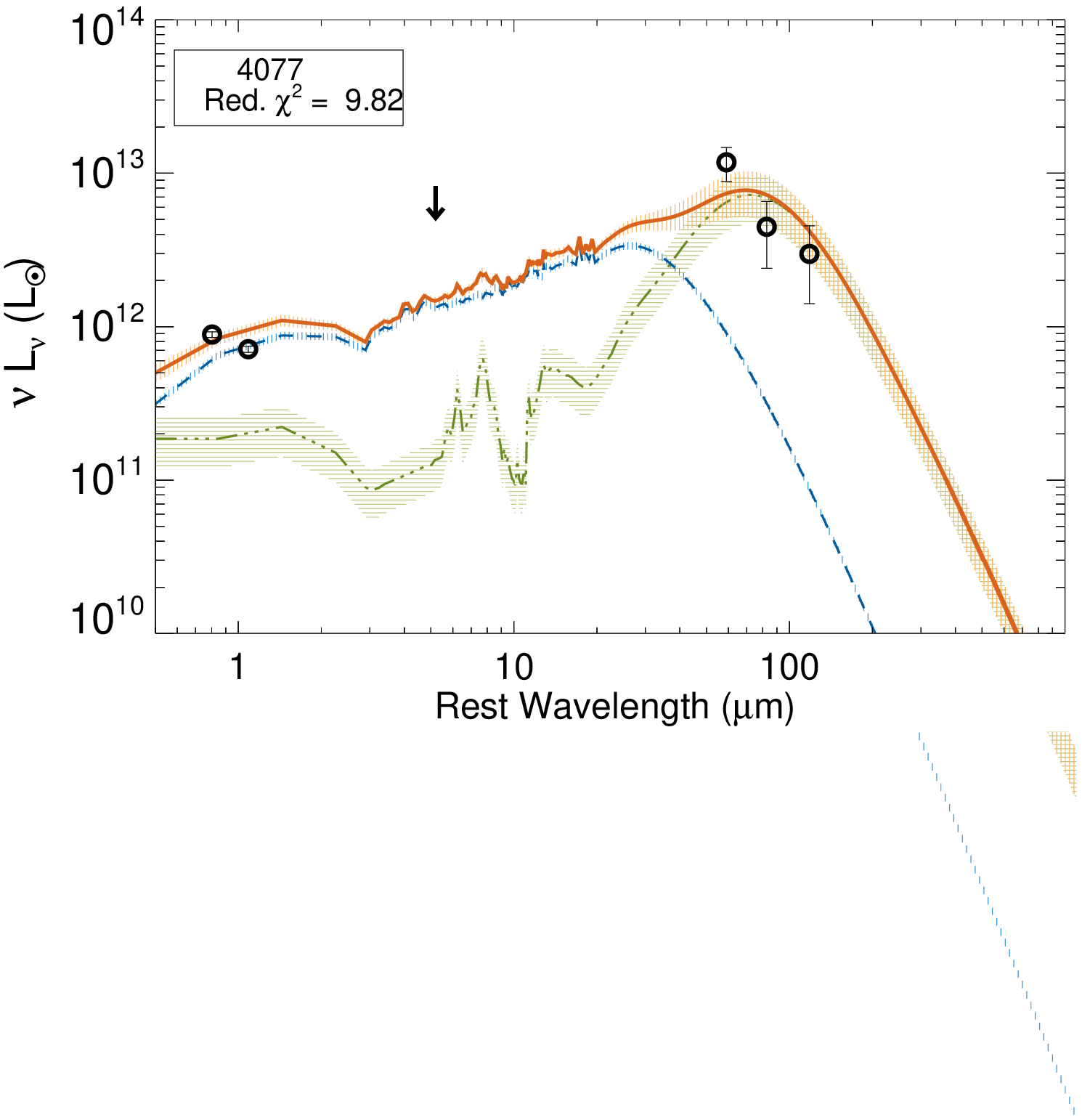}
\includegraphics[width=2.4in]{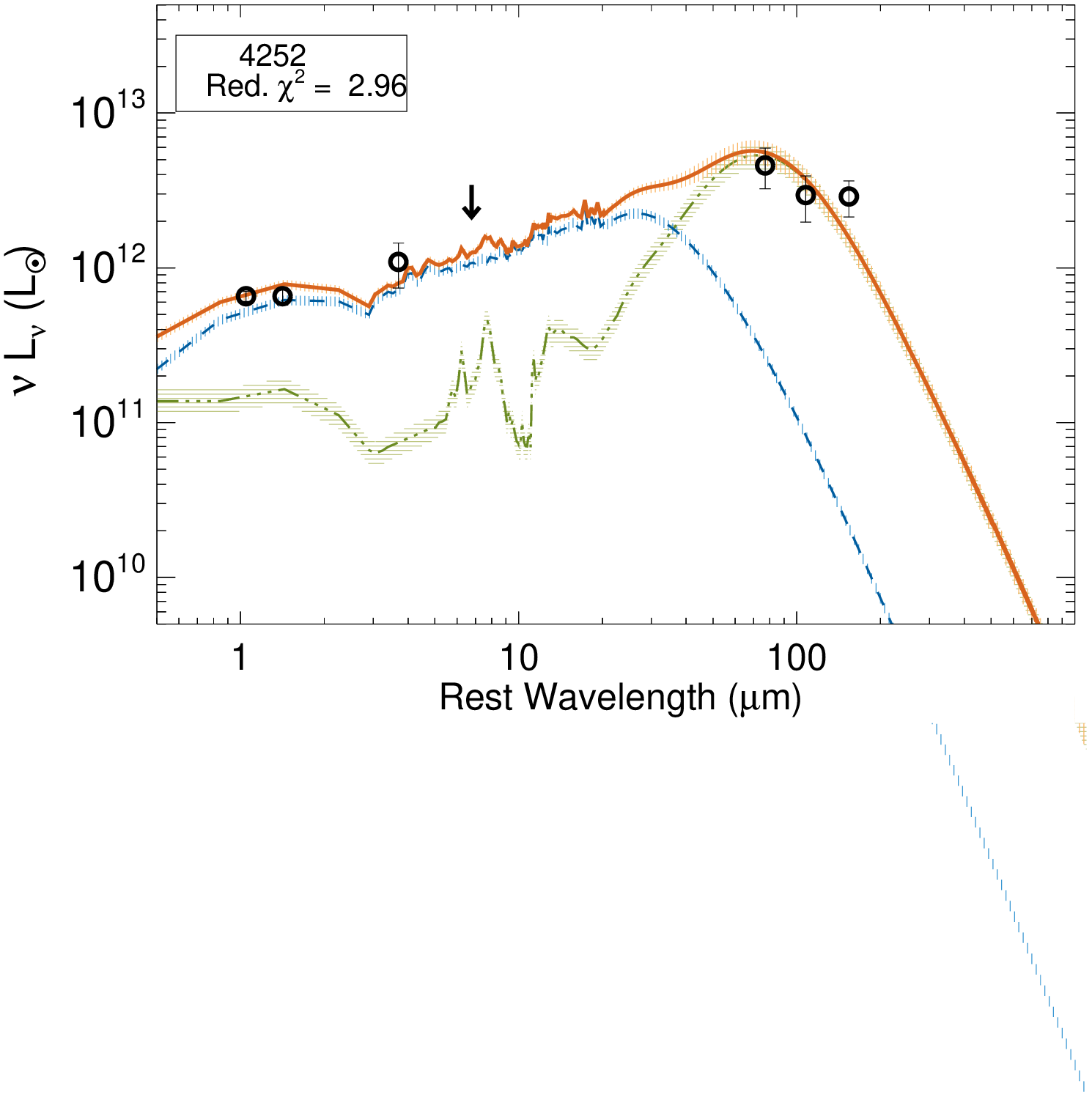}
\includegraphics[width=2.4in]{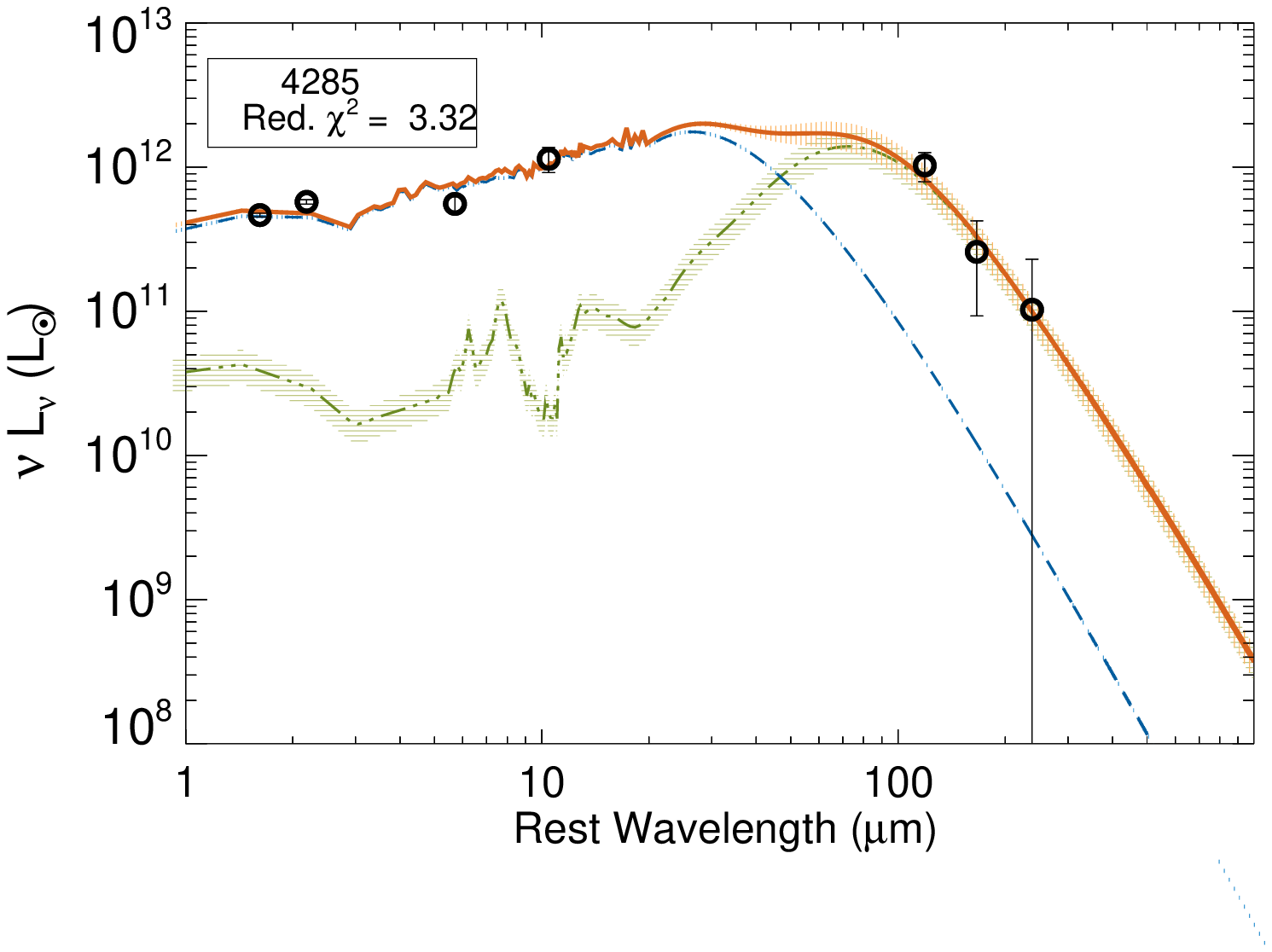}
\includegraphics[width=2.4in]{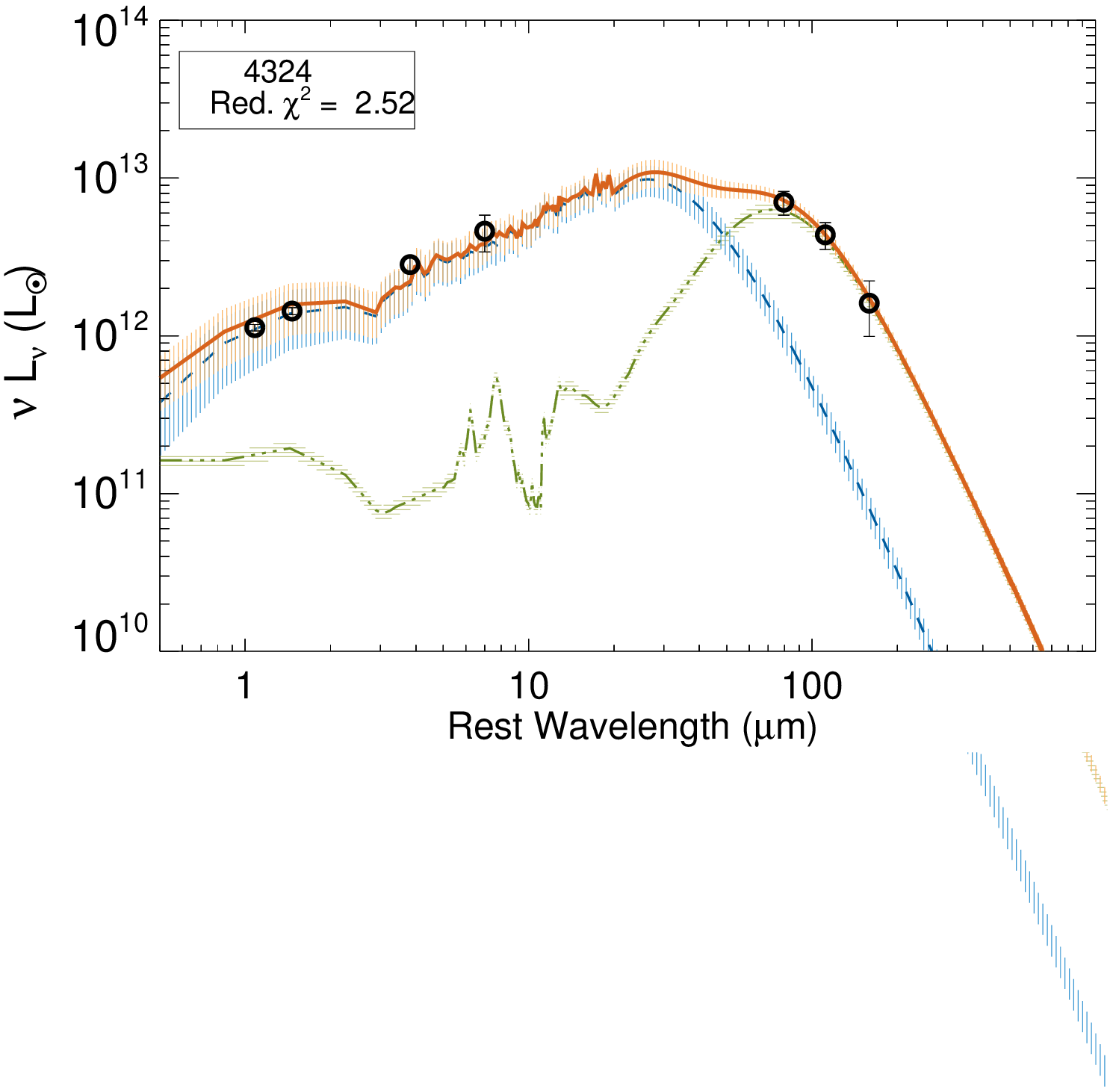}
\includegraphics[width=2.4in]{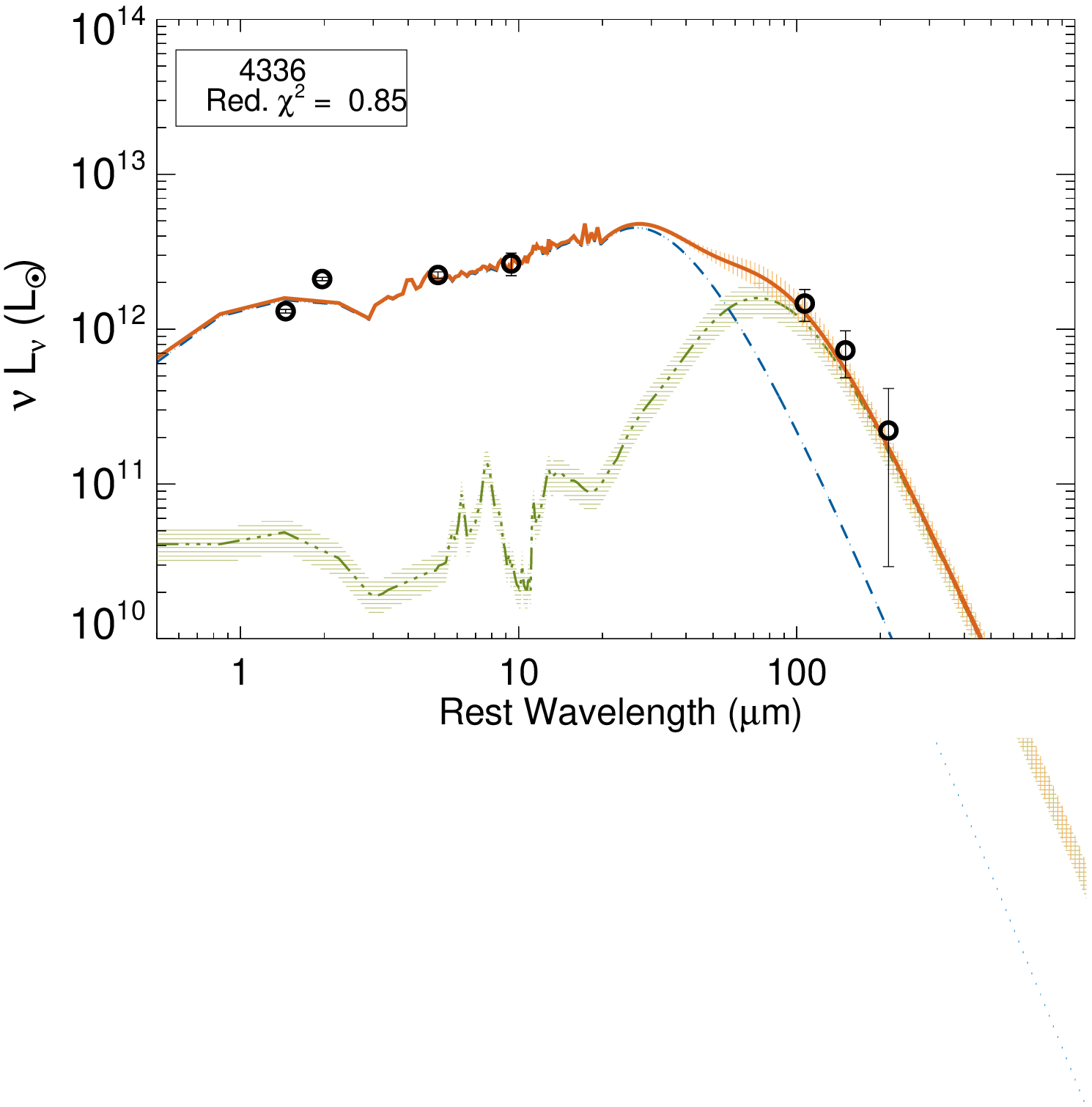}
\caption{IR decomposition for all Cold Quasars. The black circles are the WISE and {\it Herschel} photometry, the orange line is the best fit model, the blue line is the Featureless AGN template, and the green line is host galaxy. The legend of each fit lists the reduced $\chi^2$. \label{sed3fit}}
\end{figure*}

\setcounter{figure}{11} 
\begin{figure*}
\includegraphics[width=2.4in]{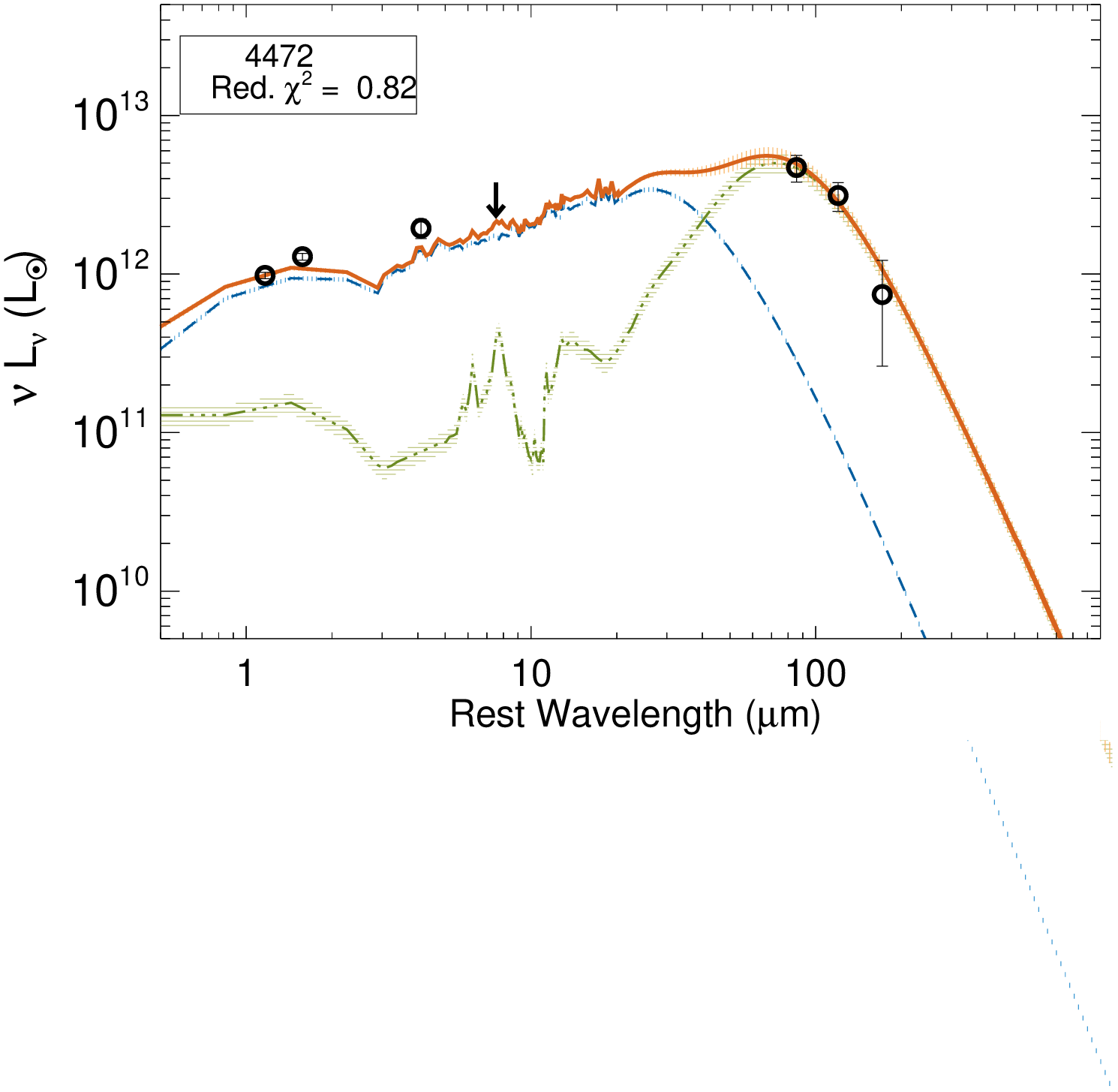}
\includegraphics[width=2.4in]{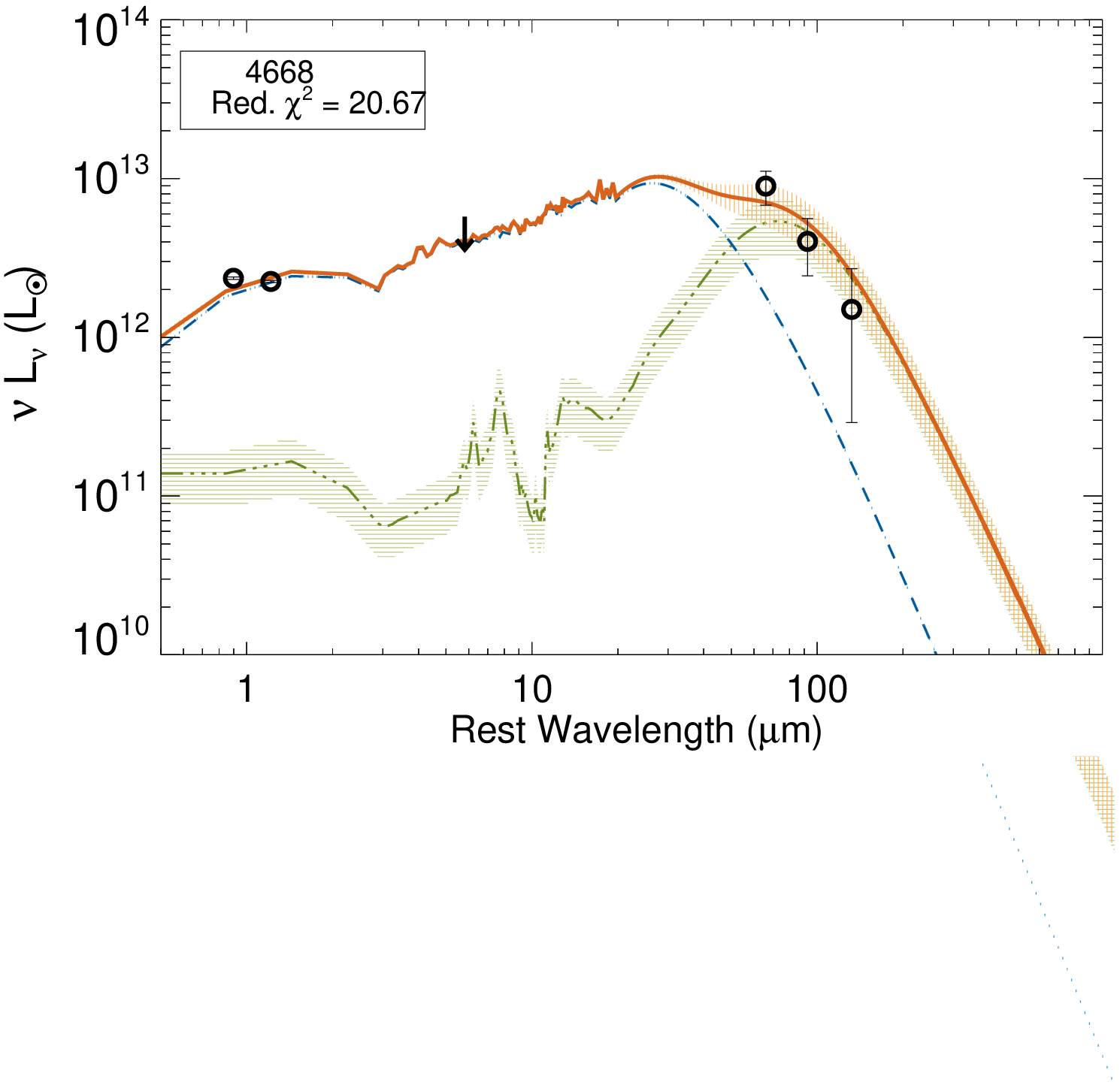}
\includegraphics[width=2.4in]{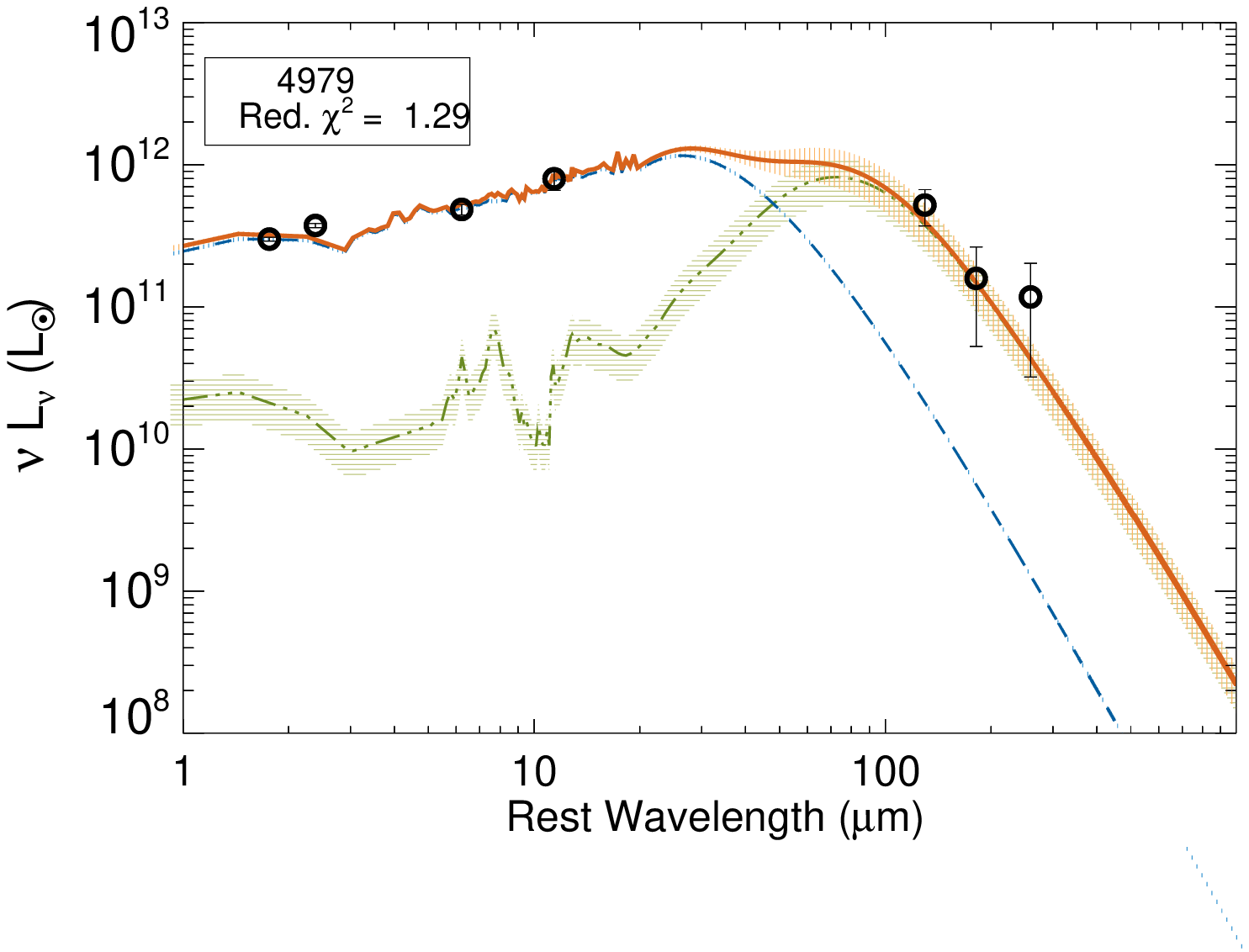}
\includegraphics[width=2.4in]{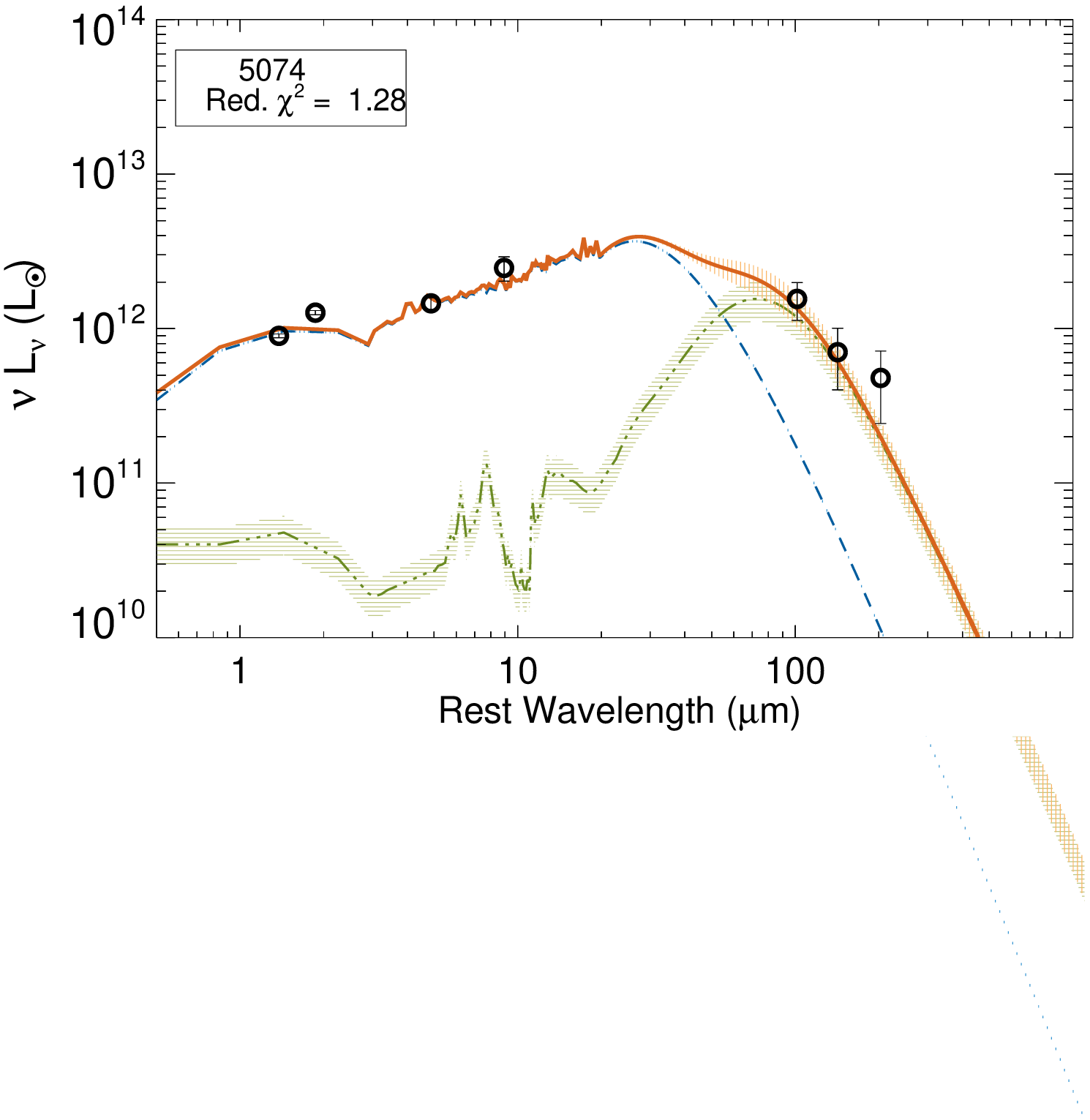}
\includegraphics[width=2.4in]{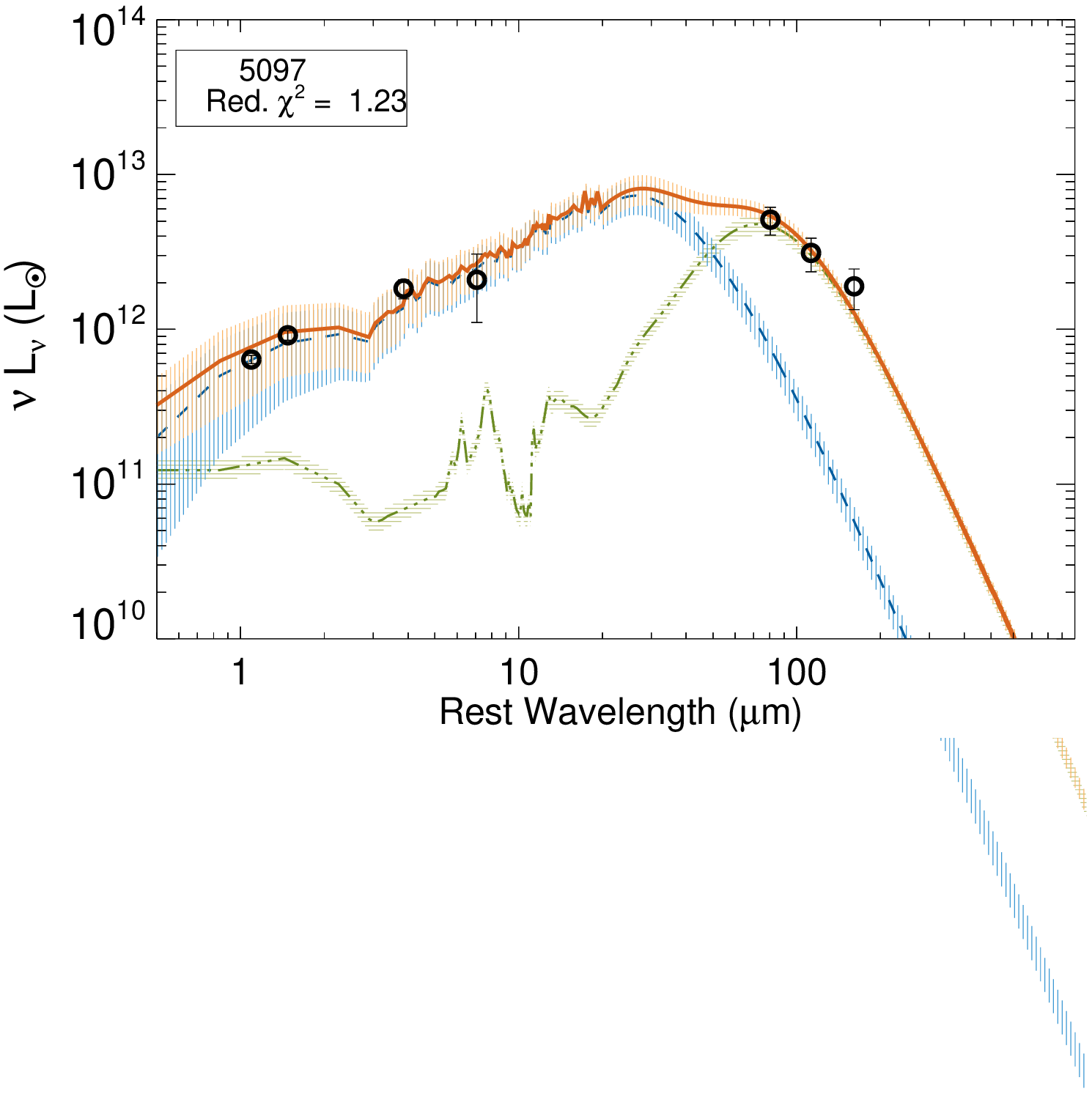}
\includegraphics[width=2.4in]{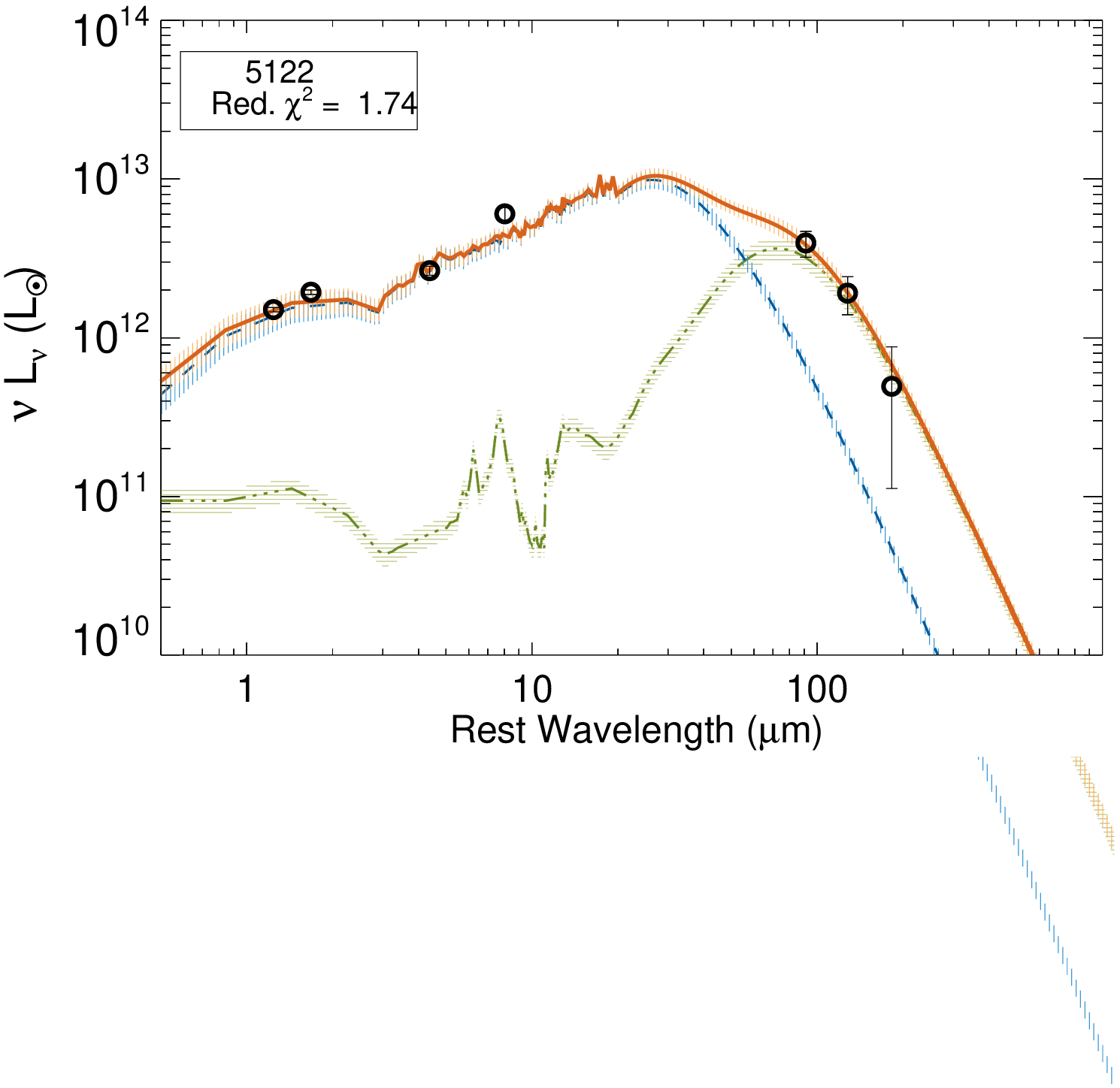}
\caption{\it (continued).}
\end{figure*}

\end{document}